\documentclass[12pt, draftclsnofoot, onecolumn]{IEEEtran}
\IEEEoverridecommandlockouts
\usepackage{cite}
\usepackage{textcomp}
\usepackage{amsmath}
\usepackage{amssymb}
\usepackage{bm}
\usepackage{graphicx}
\usepackage{xcolor}
\usepackage{stackengine}
\usepackage{mathtools}
\usepackage{subfigure}
\usepackage{bbding}
\usepackage{caption}
\usepackage{float}

\newcommand{\norm}[1]{\left\lVert#1\right\rVert}

\begin{document}
\title{Multibeam Sparse  Tiled  Planar Array for Joint Communication and Sensing}

\author{
	\IEEEauthorblockN{Hadi Alidoustaghdam$^*$, André Kokkeler, Yang Miao }\\
	\IEEEauthorblockA{Department of Electrical Engineering, University of Twente, the Netherlands\\
		\Envelope \emph{ $^*$: hadi.alidoustaghdam@utwente.nl}}
}

\maketitle

\begin{abstract}
Multibeam analog arrays have been proposed for millimeter-wave joint communication and sensing (JCAS). We study multibeam planar arrays for JCAS, providing time division duplex  communication and full-duplex sensing with steerable beams. In order to have a large aperture with a narrow beamwidth in the radiation pattern, we propose to design a sparse tiled planar array (STPA) aperture with affordable number of phase shifters. The modular tiling and sparse design of the array are non-convex optimization problems, {however, we exploit the fact that the more irregularity of the antenna array geometry, the less the side lobe level}.  We propose to first solve the optimization by the maximum entropy in the phase centers of tiles in the array; then we perform sparse subarray selection leveraging the geometry of the sunflower array. While maintaining the same spectral efficiency in the communication link as conventional uniform planar array (CUPA), the STPA improves angle of arrival estimation when the line-of-sight path is dominant, e.g., the STPA with 125 elements  distinguishes two adjacent targets with 20$^\circ$ difference in the proximity of boresight whereas CUPA cannot. Moreover, the STPA has a  40\% shorter blockage time compared to the CUPA when a blocker moves in the elevation angles. 
\end{abstract}


\section{Introduction}
In migration to 6G and connected smart networks, communication and sensing devices have to be co-designed to maintain the reliability of high data-rate connectivity and to have accurate information on the surrounding operation environment \cite{de2021convergent}. 
There are many open challenges for joint communication and sensing (JCAS), including the waveform design, the array aperture design, and the resource optimization \cite{de2021convergent,HadiSymp,zhang2021overview}. 
In this paper, our interest lies in the base station (BS) array design.
To situate both the communication and sensing functionality, we focus on enabling time division duplex (TDD) communication and a phased array radar in one array aperture.

Conventionally, the free-space radiation parameters, i.e., the directivity, half-power beamwidth and side lobe level (SLL) are the figure-of-merits for designing antenna arrays,  which is valid for the channels with dominant line-of-sight (LOS) path in communication and radar. However, in case of channels with considerable non-line-of-sight (NLOS) paths or multi-user multiple-input-multiple-output (MIMO) communication, the spectral efficiencies of communication links are the reliable performance metrics \cite{oliveri2019new,ma2021irregular}. Therefore, the radiation pattern in the BS is determined by the channel dynamics and is configured for maximizing  spectral efficiency of the communication link; it might be narrow or wide beamwidth, single or multi beams. Additionally, a phased array radar has a channel with a dominant LOS path, hence single beam with low SLL and narrow beamwidth is always beneficial for improving signal-to-noise ratio (SNR) and cross-range resolution \cite{athley2007radar}.
In regard to JCAS, a dynamic radiation pattern in BS might be desired as either communication or sensing might dictate to react to the dynamic channel conditions. To enable JCAS, the antenna selection, i.e., the synthetic change in aperture size \cite{meng2022vehicular} could be necessary. But as far as the LOS path is dominant in JCAS, which is considered in this paper, the exploitation of all antennas in the BS is beneficial for increasing the SNR and performance of JCAS. It is stated in \cite{gupta2019design} that increasing the size of aperture while preserving the same antenna numbers does not improve the directivity, therefore directivity is  not included in our study, and  we focus on designing a JCAS array with  low SLL and narrow beamwidth  in the radiation pattern while maintaining a relatively  appropriate number of radiating elements to lower the costs. 
In order to manage the number of array elements without deteriorating the performance of communication and sensing, array tiling \cite{ma2019pattern,anselmi2017irregular,dong2020low,ma2020thinned,ma2021irregular,oliveri2019new} sparse arrays \cite{vigano2011sunflower} and modular design \cite{gupta2019design}  are potential solutions.
  
 \subsection{State-of-The-Art of Large Array Design for Communication or Sensing}

\subsubsection{Array tiling} 
By grouping antennas into the so-called “tiles” and feeding each tile with one phase shifter, it guarantees a large aperture with minimum phase shifters and reduces the hardware cost.
Until now, the investigations on array tiling and the discussions on the resultant performance are mainly for communication by phased array antennas \cite{ma2019pattern,anselmi2017irregular,dong2020low,ma2020thinned} or massive MIMO communication \cite{ma2021irregular,oliveri2019new}, in which genetic algorithm, iterative convex programming or information-theoretic entropy concept are used for tiling the array. Among these solutions, authors in \cite{ma2019pattern} have proposed that two adjacent antennas can be excited by the same phase shifter hence regarded as one domino tile, and  if the phase centers of these dominoes are distributed with maximum entropy in the aperture, the radiation pattern has a low SLL.  We choose this method due  to its simplicity and manageable time cost in the optimization.

\subsubsection{Sparse modular subarray}
A narrow beamwidth in radiation pattern of a phased array requires a large array aperture size.
To decrease the complexity of large aperture, sparse modular  arrays are studied in \cite{gupta2019design} where the positions of modular uniform subarrays are optimized for desired beamwidth and SLL. When sparse arrays are considered, clustering antennas is another solution to avoid complexity of transmission lines from the RF front-end to antennas, e.g.,
in  \cite{aslan2021synthesis} and \cite{aslan2021system}, the BS antenna array is designed based on irregular clustering and sequential rotation where the optimization problem considers the inter-element spacing of elements, aperture size and modular layout design. Sparse subarrays are also proposed in non-terrestrial networks. For instance, due to a large transmission distances in satellite communications, a larger aperture is desirable to lower beamwidth; a sunflower sparse array of uniform hexagonal subarrays is designed for satellite application in \cite{vigano2009spatial,vigano2011sunflower}.

Depending on the separation \cite{9226446}  or sharing of waveforms \cite{zhang2018multibeam} between communication and sensing, the requirements of apertures for JCAS are different, e.g., number and locations of antennas or interference mitigation, but in  case of dominant LOS path, the sparse large array is beneficial for better angle of arrival (AoA) estimation \cite{athley2007radar,gupta2019design}, therefore we intend to exploit the aforementioned solutions to design a large aperture with affordable number of phase shifters  for a BS. 
We propose a sparse tiled planar array (STPA) with a shared waveform  for JCAS. Firstly, a large aperture with uniformly distributed antennas is tiled by dominoes based on the maximum entropy of tile centers in the aperture. Then the sunflower array is exploited to locate the center of subarrays and lower the number of tiles. When the tiles in subarrays are chosen, the positions of the subarrays are  optimized to lower SLL.
 \subsection{Contributions of This Paper}
As a solution for JCAS,  we propose a single STPA which is exploited for both TDD communication and sensing, besides an auxiliary aperture for radar full-duplex operation. The main contributions of this paper are:
\begin{itemize}
\item  In literature, JCAS is studied by employing conventional uniform planar arrays (CUPA)  for either downlink \cite{temiz2020dual} or uplink communication \cite{ni2021uplink}, whereas in this paper, STPA is proposed for both downlink and uplink communication at a BS since  a narrow beamwidth of the transmit beam (in downlink) or receive beam (in uplink) is required for proper AoA estimation in sensing.  The numerical results  are provided for downlink, however STPA can be adopted for uplink sensing by the method in \cite{temiz2021dual}.
\item The angular resolution of the JCAS aperture is improved by suitable adaptation of tiling and then sparse design. As it is stated in \cite{8631187}, the tiled array has the benefits of modular design and low SLL in wide scanning angles  compared to solely sparse design. The narrow beamwidth of STPA compared to CUPA ensures illuminating and receiving from only the desired angle during downlink and  uplink sensing, respectively. 
This narrow beamwidth is profitable for increasing the number of beams for multi-user communication as well. The narrow beamwidth is achieved with the cost of higher SLL compared to CUPA, however this drawback is tolerable in the channels with dominant LOS path. 
 \item Considering the large bandwidth in JCAS for 6G,  the irregularity introduced by tiling and sparse subarray geometries, compared to the conventional uniform arrays, results in a low SLL in the radiation pattern over a larger frequency band  \cite{spence2008design}. The numerical results show that STPA has a proper SLL  when the frequency is scaled up to 2.5 times  the designed frequency of operation, which results in a 42\% increase in the operational frequency of the array.
  \item The narrow beamwidth of STPA is beneficial for decreasing the blockage time since the spectral efficiency is dropped when the blocker reaches the narrower communication beam, i.e., we show in a JCAS scenario that if a tracked target blocks the communication link, the designed STPA can have a $40\%$ shorter blockage time than CUPA in elevational movements of the blocker. 
\end{itemize}
The following notations are used in this paper:
$(.)^T$ and $(.)^*$ denote the transpose and conjugate transpose operator, respectively. $\norm{.}_F$ is the Frobenius norm.
 \begin{figure}[ht]
	\centering
	\includegraphics[width=0.85\textwidth]{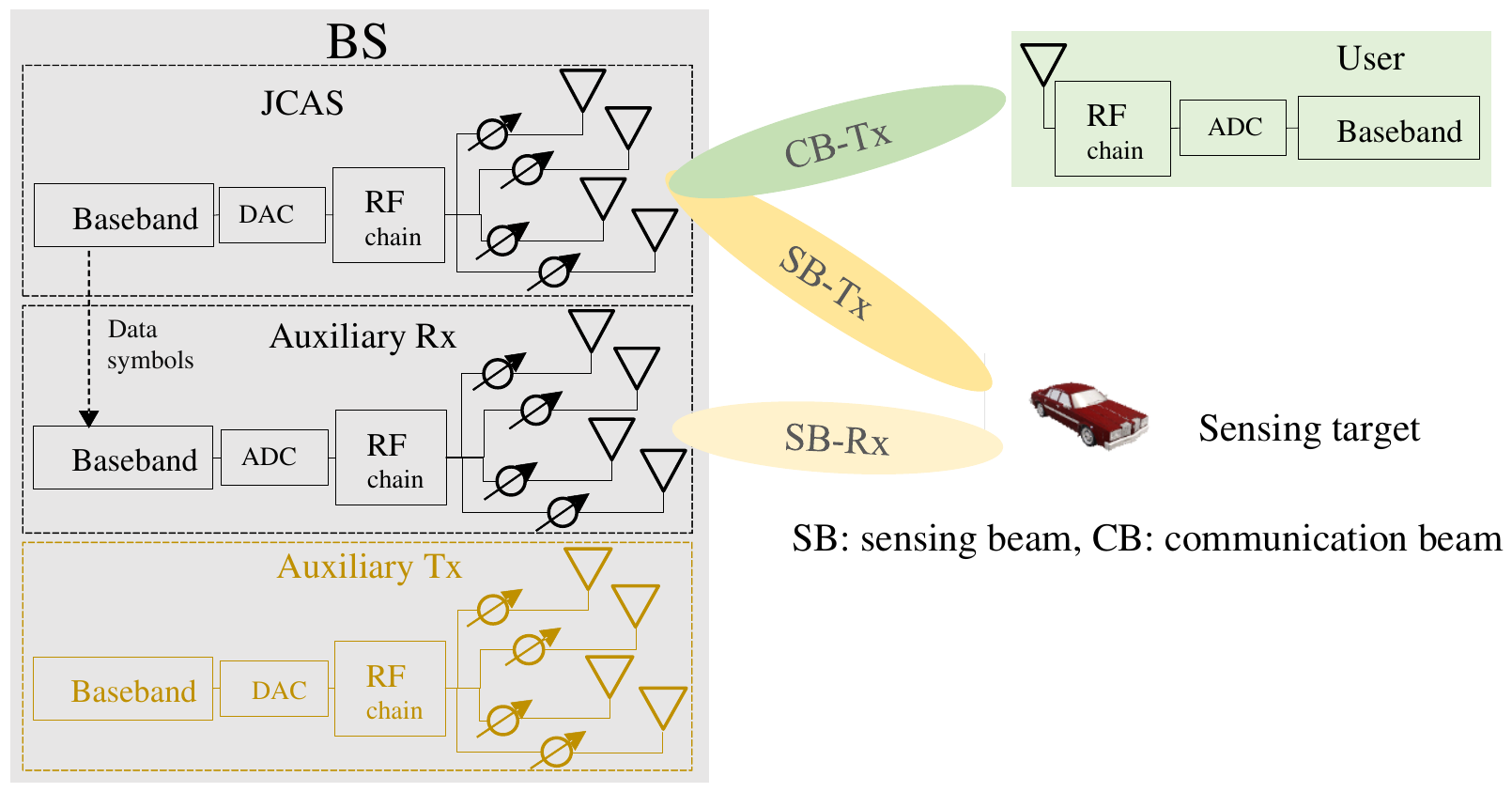}
		\centering
	\caption{The configuration for JCAS in BS while operating in downlink communication.} 
	\label{JCAS_Config}
  \end{figure}
\section{Proposed Design Methodology}
In this section, the structure of a BS with an analog beamforming array for JCAS is studied. After that, a realization of STPA is proposed via a flowchart and finally the metrics adopted for JCAS performance are introduced.  
\subsection{Problem Formulation}
Let us consider a TDD base station based on fully connected analog array as in
Fig.~\ref{JCAS_Config}, where a JCAS aperture with $N$ antennas is supposed to communicate with a single-antenna user and illuminate the target via two beams. This aperture serves as the transmitter during downlink communication and as the receiver during uplink communication. Without loss of generality, we study the performance of only one RF chain of a fully connected analog array, since other RF chains are also connected to the same antennas.
Due to long communication frames, full-duplex radar sensing functionality is mandatory for perpetual sensing of environment, therefore  auxiliary apertures with $N_{aux}$ antennas are needed to serve as radar receiver during downlink communication, and as radar transmitter during uplink communication\footnote{The reason for transmit antennas at the uplink is that sensing solely by the communication signal has the following problems, i.e., the power resources of communication users are limited so is the SNR for sensing,  these users have also beamforming toward BS, therefore the environment is not illuminated for sensing, besides the sensing environment is dependent on the location of user which could be unknown, therefore transmit antennas at BS during uplink communication are beneficial for sensing
 \cite{temiz2021dual}.}.
 The signal processing for downlink and uplink JCAS are different, i.e., reflected signals from sensing target and uplink communication signal from the user interfere at JCAS aperture which operates as the receiver, hence they need to be separated, whereas downlink sensing does not need such separation since the auxiliary Rx only receives the reflection from the target. In this paper, we choose downlink communication and sensing to demonstrate the performance of STPA. This performance is representative to the uplink communication and sensing, since the narrow beamwidth of the JCAS aperture as a receiver is beneficial for the sensing, and the spectral efficiency of uplink communication with the STPA is similar to the CUPA.  
 \par
 Radar has a two-way radiation pattern, and an ideal radar radiation pattern should be narrow in both transmit and receive pattern for optimal power usage, hence we study the challenging part where both communication and sensing are considered in a single JCAS aperture, the auxiliary arrays can be designed accordingly. 
One of the features of millimeter wave (mmWave) communication is the large number of antennas, which provides larger gain towards the user. Besides, the antennas in the JCAS aperture are used as transmitter for downlink sensing,
therefore the transmit beam for sensing is already narrow
and the auxiliary Rx can have less antenna numbers to decrease implementation costs, hence we assume $N\gg N_{aux}$. 
Therefore, the tiling is targeted only to the $N$ antennas at the JCAS aperture, and we simply choose $N_{aux}=1$ and focus on the JCAS aperture for the performance study of STPA. We use communication signals with orthogonal frequency division multiplexing (OFDM) waveforms to communicate and sense.
  \begin{figure}[H]
	\centering
	\includegraphics[width=1\textwidth]{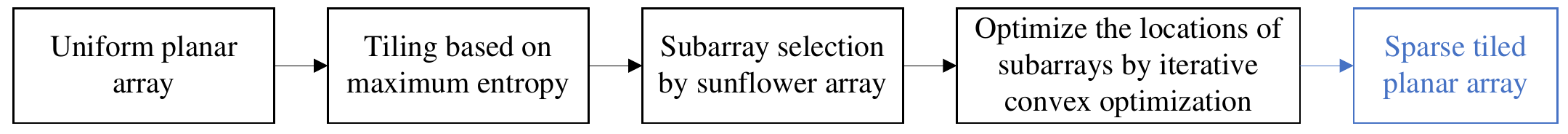}
	\caption{The flowchart of designing STPA.}
	\label{FlowchartDesign}
  \end{figure}
\subsection{Realization of STPA}
\begin{figure}[ht]
	\centering
	\includegraphics[width=0.4\textwidth]{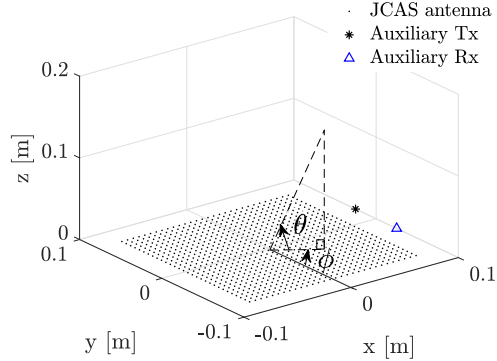}
	\caption{The initial configuration of antennas at the BS.} 
	\label{Geo1}
  \end{figure}

 Note that the inherent problems of  tiling of the JCAS aperture for the least SLL and a sparse aperture for the best AoA estimation are non-convex optimizations which are difficult to solve, therefore the optimization has to be based on the foreknowledge, hence local optimization is employed, as it is shown in the flowchart of Fig.~\ref{FlowchartDesign}. The main idea of this design is based on two scales of irregularity, namely the geometry of  subarrays (local scale), and the phase centers of tiles in the whole aperture (global scale).  Initially, a large CUPA is tiled for the minimum SLLs in the radiation pattern.
 The tiling problem is formulated based on maximum entropy  in the distribution of the phase centers
of the tiles in the array.
In the JCAS aperture shown in Fig.~\ref{Geo1}, assume $N_\text{sub}$ antennas are grouped in  tiles and fed by power dividers. The center of mass of such a tile could be considered as the center of phase for the radiation from the whole tile.
The irregularity in the distributions of these phase centers is measured
with an information-theoretic entropy-based objective function, i.e., if the entropy of phase centers at each row and column of the aperture is maximized, then the resulting phased array aperture has a desired SLL while scanning. To obtain more mathematical insight into  tiling based on the maximum entropy, refer to \cite{ma2020high}. The resultant tiled array  can still be thinned by grouping some tiles in subarrays and removing others.  A subarray is determined by a circle circumventing its tiles, the centers of these circles (subarrays) are obtained by the geometry of the sunflower array  since the positions of elements are controllable, although it is an aperiodic sparse array (see appendix A). The radius of a circle is a thinning parameter for the aperture, which can be tuned to reach the desired SLL with minimum number of tiles. It is assumed that the phase centers of tiles are still distributed with a maximum entropy after thinning of aperture, but the inter-distances among subarrays increase the SLL, however the resultant empty space among subarrays can be exploited to optimize the location of subarrays efficiently to decrease the SLL (see appendix B).  Once the JCAS aperture with the minimum  number of phase shifters, the desired SLL and beamwidth is achieved, its performance is compared to CUPA.

\subsubsection{Expanded Beam Pattern}
 In order to assure that the designed STPA bears the desired SLL and beamwidth in various scanning scenarios, the expanded beam pattern (EBP) is employed, which facilitates studying SLLs for all scanning angles in a single radiation pattern.
The radiation field of an array with isotropic radiating  elements  in $(u,v)$-space can be obtained by:
\begin{equation}
f(u,v)=\frac{1}{N}\sum_{n=1}^{N}\alpha_n e^{j\Omega_n} e^{jk(ux_n+vy_n)},u=\cos\theta\cos\phi,v=\cos\theta\sin\phi
  \label{rad0}
\end{equation}
where $N$ is the number of  radiating elements, $\alpha_n$ and  $\Omega_n$ are the amplitude and phase of excitation, respectively. $k=2\pi/\lambda$ is the wave-number at the wavelength $\lambda$, $x_n$ and $y_n$ are the positions of antennas in  Cartesian coordinates. $\theta$ is the elevation angle measured from $xy$-plane and $\phi$ is  the azimuth angle measured from $x$-axis. 
The radiation field in \eqref{rad0} can scan toward the angle $(u_s,v_s)$ by the proper weighting of amplitudes and phases. In case of conventional beamforming, which is based on only phase excitation in the elements, the weights are:
\begin{equation*}
    \alpha_n=1,\Omega_n=-jk(u_sx_n+v_sy_n),
\end{equation*}
 then the radiation pattern is:
 \begin{equation}
  F(u,v;u_s,v_s)=\frac{1}{N^2}\Bigg|\sum_{n=1}^{N}e^{jk\big((u-u_s)x_n+(v-v_s)y_n\big)}\Bigg|^2
  \label{rad1}
\end{equation}
  In order to simplify studying the above radiation pattern for all $(u_s,v_s)$s, the EBP is defined as \cite{gupta2019design}:
  \begin{equation*}
      F_{\zeta}(\tilde{u},\tilde{v})=
      \frac{1}{N^2}\Bigg|\sum_{n=1}^{N}e^{jk\zeta(\tilde{u}x_n+\tilde{v}y_n)}\Bigg|^2, \zeta=1+\cos(\theta_{min})
  \end{equation*}
 where $\tilde{u}=u-u_s$, $\tilde{v}=v-v_s$. The scanning of beam in the whole $0^\circ\leq\theta\leq 90^\circ$ without grating lobes is not practical, therefore the desired angle of view is $\theta_{min}\leq \theta \leq 90^\circ$ where $\theta_{min}$ is the minimum of angle $\theta$. The  SLL for $F_{\zeta}(\tilde{u},\tilde{v})$ is studied to ensure an appropriate SLL in all scanning angles.
 \begin{figure*}[tb]
	\centering
	\includegraphics[width=0.9\textwidth]{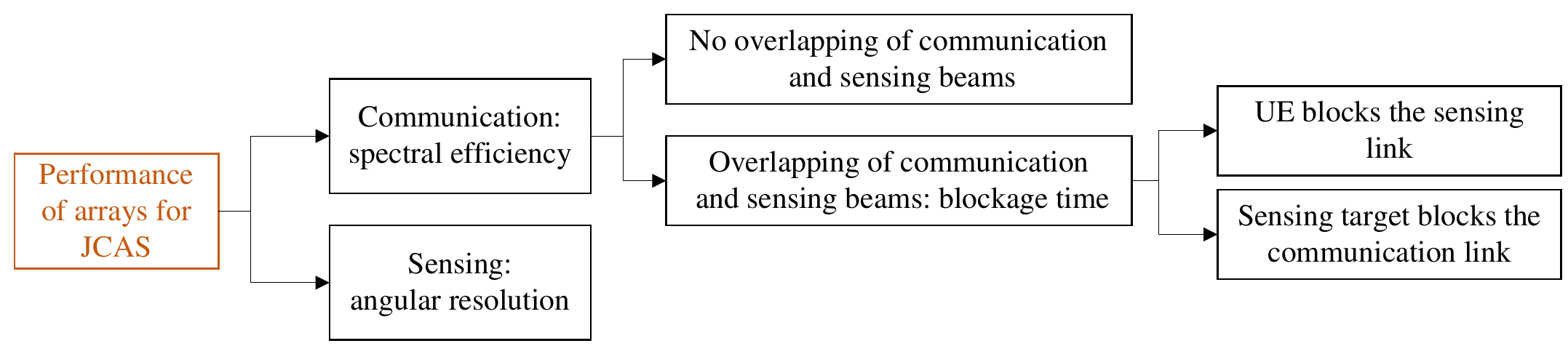}
	\caption{The flowchart of performance measurement.}
	\label{FlowchartPM}
  \end{figure*}
 \begin{figure*}[tb]
	\centering
	\includegraphics[width=0.9\textwidth]{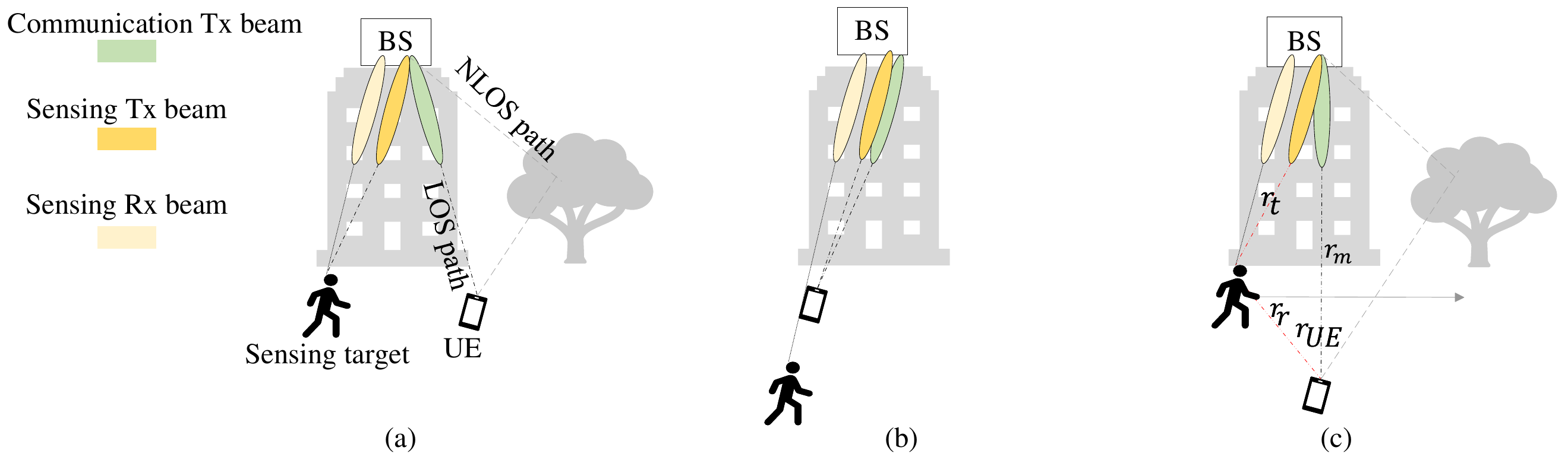}
	\caption{The scenarios for JCAS in downlink communication, (a) without overlapping of communication and sensing beams,  (b) with overlapping of communication and sensing beams, when the UE blocks the sensing link and (c) forthcoming overlapping of communication and sensing beams, when the sensing target can block the communication link.}
	\label{CP}
	\end{figure*}
\subsection{Performance Metrics}
The  performance metrics for communication and sensing are spectral efficiency and angular resolution, respectively, as  it is shown in  the flowchart of Fig.~\ref{FlowchartPM}. The communication performance (CP) includes cases  without and with overlapping of beams when the UE blocks the sensing link and also when the sensing target blocks the communication link. 
The overlapping occurs when:
\begin{equation}
(u_c-u_s)^2+(v_c-v_s)^2<r_b^2
\end{equation}
where $(u_c,v_c)$ and  $(u_s,v_s)$ are the angles of communication and sensing beams, respectively, and $r_b$ is the radius of the beamwidth  in $(u,v)$-space.
In pursuit of simplicity, we suppose the user  has no reflection of electromagnetic waves and all the energy transmitted in the communication beam of the BS is captured by the user.
\subsubsection{CP without overlapping communication and sensing beams}
Let us consider the scenario in Fig.~\ref{CP}(a). A geometric model for both communication and sensing including $L$ multipath signals with the angle of departure $(\theta_{T,l},\phi_{T,l})$ and AoA $(\theta_{R,l},\phi_{R,l})$ is assumed, then the time varying channel  $\mathbf{H}\in \mathbb{C}^{N\times N_R}$ would be:
\begin{equation}
\begin{split}
   &\mathbf{H}=\sum_{l=1}^{L}b_l\delta(t-\tau_l)e^{j2\pi f_{D,l}t}\mathbf{a}_{T}(\theta_{T,l},\phi_{T,l}) \mathbf{a}^T_{R}(\theta_{R,l},\phi_{R,l})\\
    \end{split}
     \label{Channel}
\end{equation}
where $b_l$ is the complex value modelling signal attenuation, $\tau_l$ is the propagation delay and $f_{D,l}$ is the Doppler frequency. $\mathbf{a}_{T}$ and $\mathbf{a}_{R}$ are the steering vectors for transmit and receive arrays. The channel \eqref{Channel} might also include the LOS path for communication. The multibeam weighting vector at the JCAS aperture is \cite{zhang2018multibeam}:
\begin{equation}
\mathbf{w}_{JCAS}=\sqrt{\rho}\mathbf{w}_{c}+\sqrt{(1-\rho)}\mathbf{w}_{s}
\label{w_JCAS}
\end{equation}
where $0\leq\rho\leq1$ is the power allocation factor between beams which equals to 0.5 in this paper, $\mathbf{w}_{c}\in \mathbb{C}^{N\times 1}$ is the weighting vector for communication and  $\mathbf{w}_{s}\in\mathbb{C}^{N\times 1}$  denotes the weighting vector for sensing to a desired direction $(\theta_s,\phi_s)$. For simplicity, $\mathbf{w}_{c}$ is considered as the right singular vector of the channel corresponding to the maximum singular value of $\mathbf{H}$ with $\norm{\mathbf{w}_{c}}_F^2=1$ whereas  $\mathbf{w}_{s}=\mathbf{a}_T^*(\theta_s,\phi_s)/\norm{\mathbf{a}_T(\theta_s,\phi_s)}_F$. 
In this paper, the performance metric for communication is spectral efficiency, and when Gaussian symbols are transmitted in a mmWave channel,  the spectral efficiency is \cite{el2014spatially}:
\begin{equation}
 R=\text{log}_2\text{det}\Big(1+\frac{\mathbf{H}^*\mathbf{w}_{JCAS}^*\mathbf{w}_{JCAS}\mathbf{H}}{\sigma_n^2}\Big)
    \label{R}
\end{equation}
where $\sigma_n^2$ denotes the variance of a zero-mean Gaussian noise.
\subsubsection{CP with overlapping beams when the UE blocks the sensing link}
When the UE blocks the sensing link, the two beams are united and the whole energy is captured by the UE, then the spectral efficiency is increased. This happens either in the scanning mode of sensing or the tracking mode with  the range of target being larger than the range of UE as in Fig.~\ref{CP}(b).
The model for this case, due to its simplicity, is not considered here and  will only be shown via simulation.
\subsubsection{CP with overlapping beams when the sensing target blocks the communication link}
Let us study the overlapping when the range of the target is smaller than the range of a single-antenna UE,  as in Fig.~\ref{CP}(c). It is supposed that the target occupies only one angle step, is tracked, and the sensing beam is aligned with the angle of the target. When this target moves toward the communication link, the beams are overlapping,  the LOS link is vanished and  blockage occurs. In order to model the blockage, the following channel model is considered with and without blockage \cite{bhardwaj2021geometrical,RizqiGuard}:
\begin{equation}
\begin{split}
&\mathbf{H} =\begin{cases}
      \mathbf{H}_{LOS}+ \mathbf{H}_{NLOS}+\mathbf{H}_{target,1} &  \text{ if no overlapping beams,}
\\
      \mathbf{H}_{NLOS}+\mathbf{H}_{target,2} &  \text{ if beams overlapping,}
    \end{cases} \\
    \end{split}
    \label{BlockedChannel}
\end{equation}
where:
\begin{equation}
\begin{split}
&
\mathbf{H}_{LOS}=\mathbf{a}_T(\theta_{UE},\phi_{UE})\frac{\lambda}{4\pi r_{UE}}e^{-j\frac{2\pi}{\lambda }r_{UE}}, \\
&
\mathbf{H}_{target,s}=\mathbf{a}_T(\theta_{t},\phi_{t})\frac{\lambda}{4\pi d_{t}}\Gamma_{t,s} e^{-j\frac{2\pi}{\lambda }d_{t}} ,
s\in \{1,2\}, d_t=r_t+r_r,
\end{split}
\end{equation}
where  $\mathbf{H}_{NLOS}$ is the same channel \eqref{Channel} as described in section C.1,   $\mathbf{H}_{target,1}$ and  $\mathbf{H}_{target,2}$ denote the channel of BS-target-UE before or after the blockage and during the blockage, respectively. $r_t$ and $r_r$ are the distance between the target and the BS, and the target and the UE, respectively.
 Let us suppose a human as the target, therefore  
$\Gamma_{t,1}$ which models the reflection and scattering is calculated by \cite{bhardwaj2021geometrical}: 
\begin{equation}
\Gamma_{t,1} (\delta)=\frac{\epsilon_b\sin\delta-\sqrt{\epsilon_b-\cos^2\delta}}{\epsilon_b\sin\delta+\sqrt{\epsilon_b-\cos^2 \delta}}+\mathcal{CN}(0,\sigma_b^2)
\label{Gamma_t1}
\end{equation}
where $\delta$ is the incident angle on the body, $\epsilon_b$ is the complex permittivity of the body and a zero-mean Gaussian random variable with a standard deviation of $\sigma_b$ models the non-flat body and swinging limbs.
Besides, when the human blocks the LOS link, the result is  $20$ dB attenuation of power at the receiver UE \cite{virk2019modeling}, therefore $\Gamma_{t,2}$ which denotes the diffraction coefficient,  simply equals to $0.1$ at the operation frequency of $28$ GHz. 
We assume the target moves  in a line perpendicular to the LOS path where  $\phi_t=\phi_{UE}$ and only the angle $\theta_t$ varies,   and its Doppler effects on $\mathbf{H}_{target,s}$ are negligible. The channel modeled in $\eqref{BlockedChannel}$ is used for calculating the spectral efficiency.

\subsection{Sensing performance of JCAS aperture}
The main sensing feature of the designed JCAS BS in Fig.~\ref{CP}(a) is the angular resolution due to its narrow beamwidth, which will be demonstrated by two angularly adjacent targets. The sensing here is based on the phased array where at each angle, the range and velocity are estimated.  Similar to \cite{zhang2018multibeam}, the same OFDM  waveform is used for JCAS and sensing should be performed during multiple sets of symbols; $N_d$ symbols are required for the coherent processing for Doppler estimation at each angle, where $N_e$ and $N_a$ are the number of elevation and azimuth angles, respectively, resulting in $N_eN_a$ angles for full sensing of space. 
Here, we suppose that self-interference is already  cancelled by a combination of passive \cite{nuss2017mimo} and active methods \cite{temiz2021dual,barneto2019full}, therefore the received signal at the auxiliary Rx is simplified by:
\begin{equation}
\tilde{y}_{n,n_a,k}=\sum_{l=1}^{L}b_l \mathbf{w}^T_{JCAS}\mathbf{a}_T(\theta_{T,l},\phi_{T,l})e^{-j2\pi n\tau_l f_0}e^{j2\pi f_{D,l}(kT_s+(n_a-1)T_f)}+\tilde{z}_n/\tilde{s}_n\\
     \label{y}
\end{equation}
where $\tilde{y}_{n,n_a,k}$ is the received signal at $n$th subcarrier when sensing the angle indexed by $(n_a,k)$ where $k=(n_e-1)N_d+n_d$, $1\leq n_e\leq N_e$ and $1\leq n_d\leq N_d$. While $T_s$ is the symbol duration, $f_0=1/T_{s}$. $\tilde{z}_n$ is the noise sample and $\tilde{s}_n$ is the baseband signal for subcarrier $n$.
It is assumed that the communication beam is constant during $T_f$, the time of one communication packet.
At each scanning angle, $\mathbf{w}_s$ and  consequently  $\mathbf{w}_{JCAS}$ in \eqref{w_JCAS} are calculated, then the
 inverse discrete Fourier transform (DFT) and DFT are applied to the received signal  $\tilde{y}_{n,n_a,k}$ to extract the range and Doppler of targets, respectively.
 \begin{figure}[H]
	\subfigure[]
	{\includegraphics[width=0.43\linewidth]{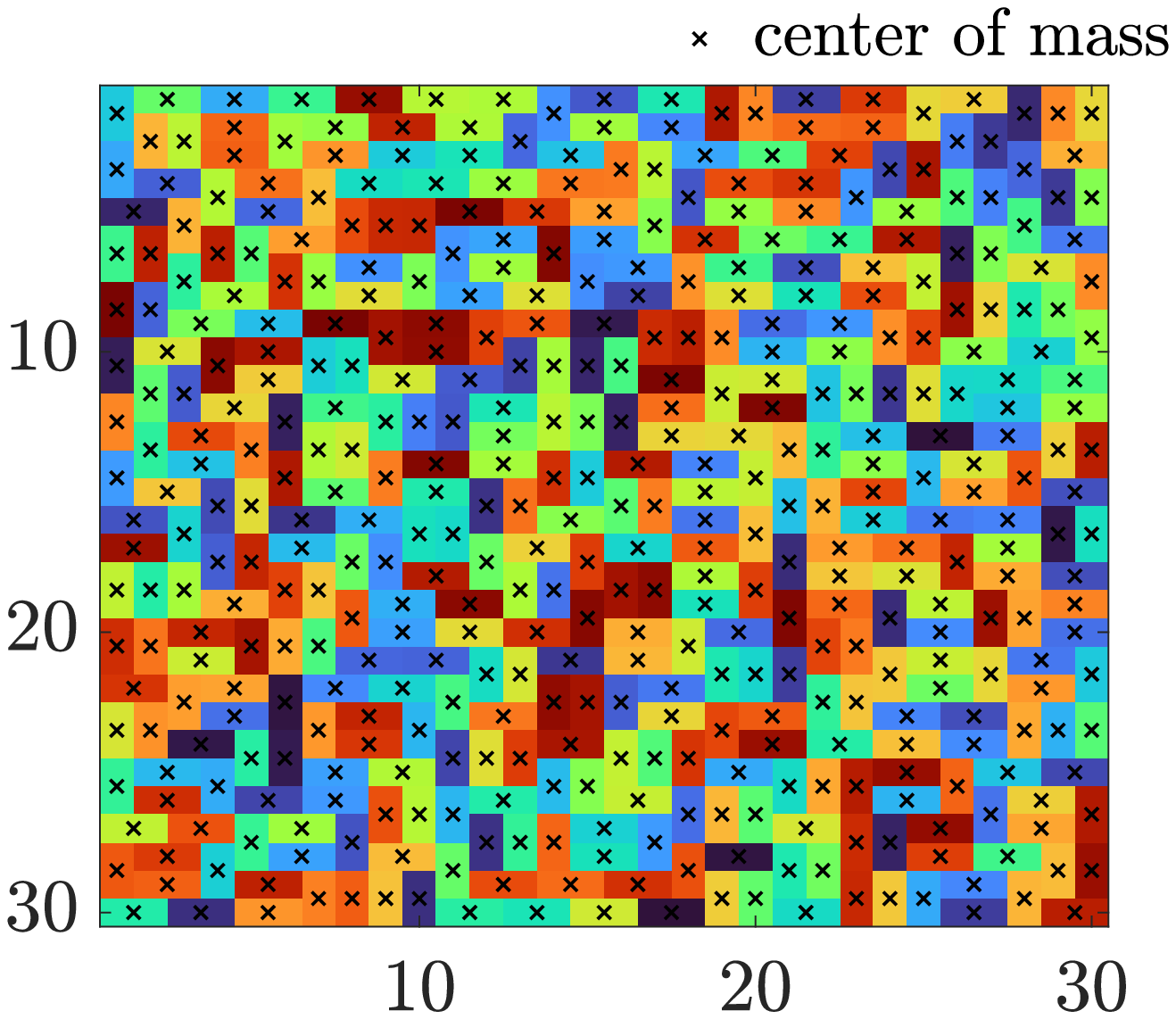}}
	\subfigure[]
   {\includegraphics[width=0.5\linewidth]{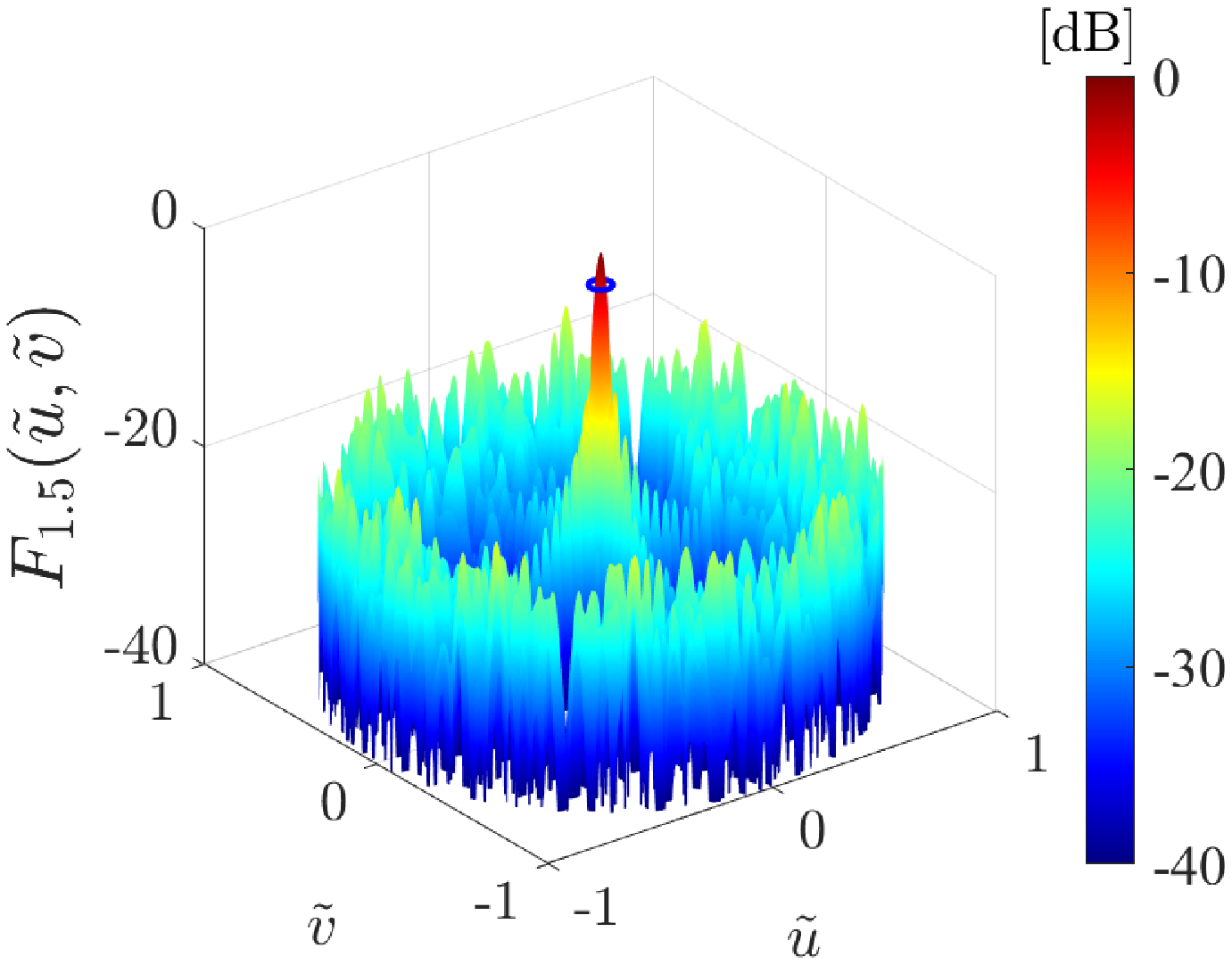}}
	\centering
	\caption{ (a) The tiled array with dominoes (the colors are only for differentiating the tiles), (b) its EBP with the blue circle with radius $\tilde{r}_b=0.043$ denotes the 3 dB-beamwidth of the main beam in $(\tilde{u},\tilde{v})$-space, then the beamwidth in $(u,v)$-space is $r_b=\zeta \tilde{r}_b=0.065$.  }
	\label{tiles}
\end{figure}
\begin{figure}[H]
    \centering
    \includegraphics[width=0.5\linewidth]{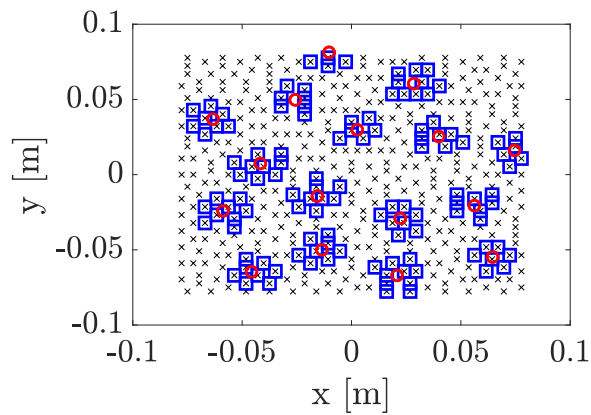}
    \caption{The subarrays chosen based on the sunflower array, where the crosses denote the phase centers of the dominoes, red circles show the centers of the subarrays, and the blue squares demonstrate the chosen dominoes in each subarray. }
    \label{SFA}
\end{figure}
\section{Numerical Analysis}
\subsection{JCAS Array Aperture – A Design Example}
\begin{figure}[H]
	\subfigure[]
	{\includegraphics[width=0.4\linewidth]{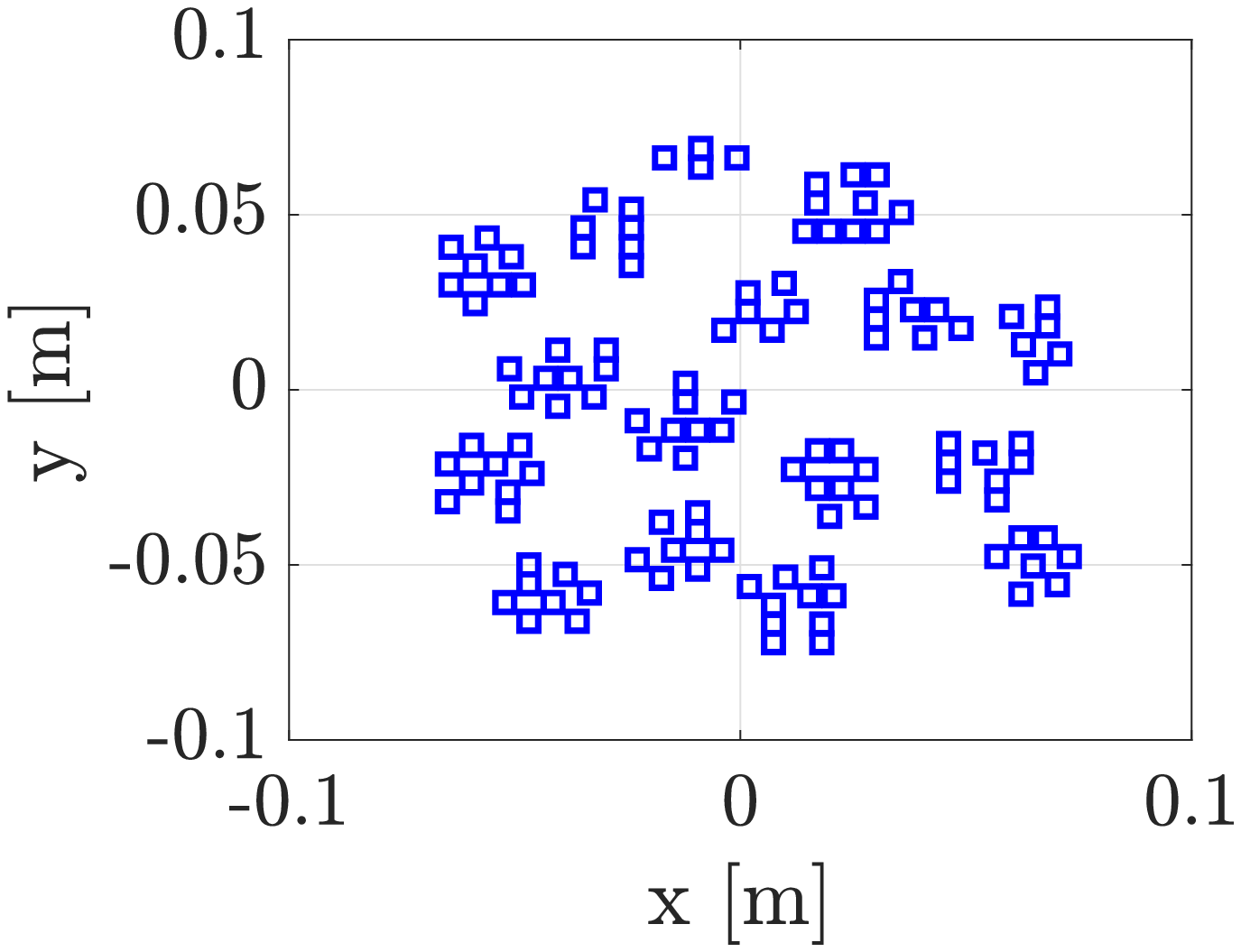}}
	\subfigure[]
	{\includegraphics[width=0.4\linewidth]{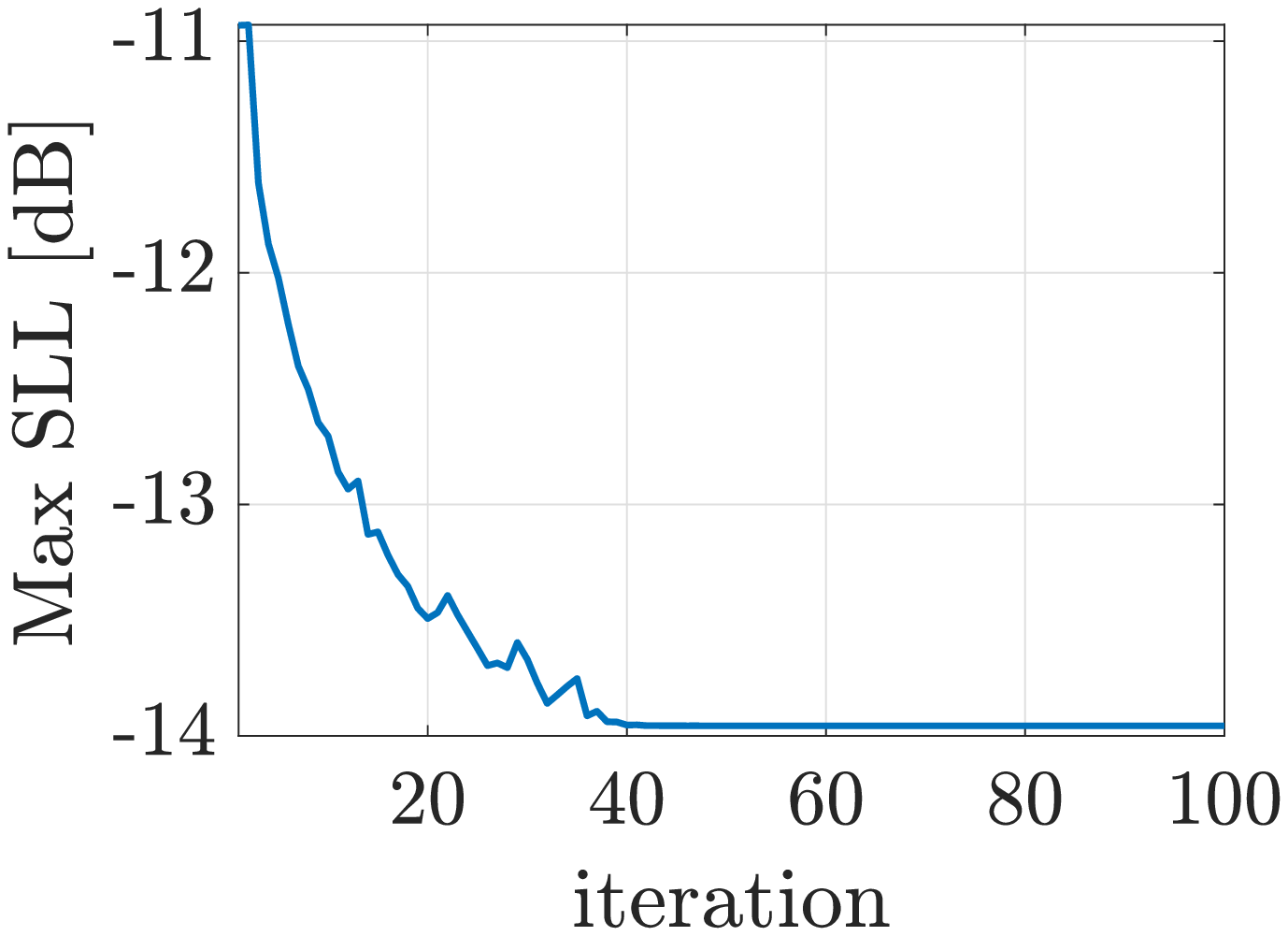}}
	\subfigure[]
	{\includegraphics[width=0.4\linewidth]{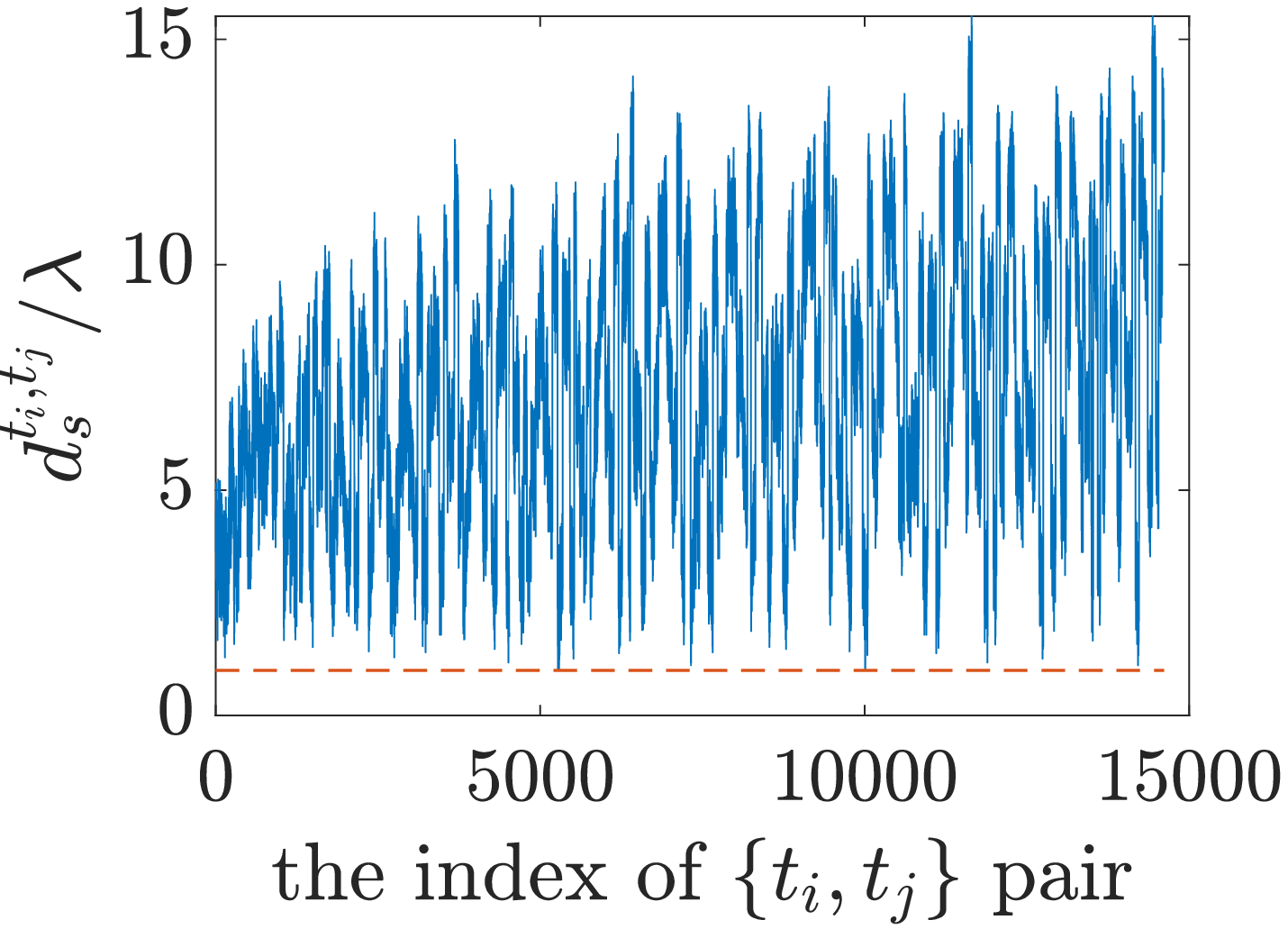}}
	\subfigure[]
	{\includegraphics[width=0.4\linewidth]{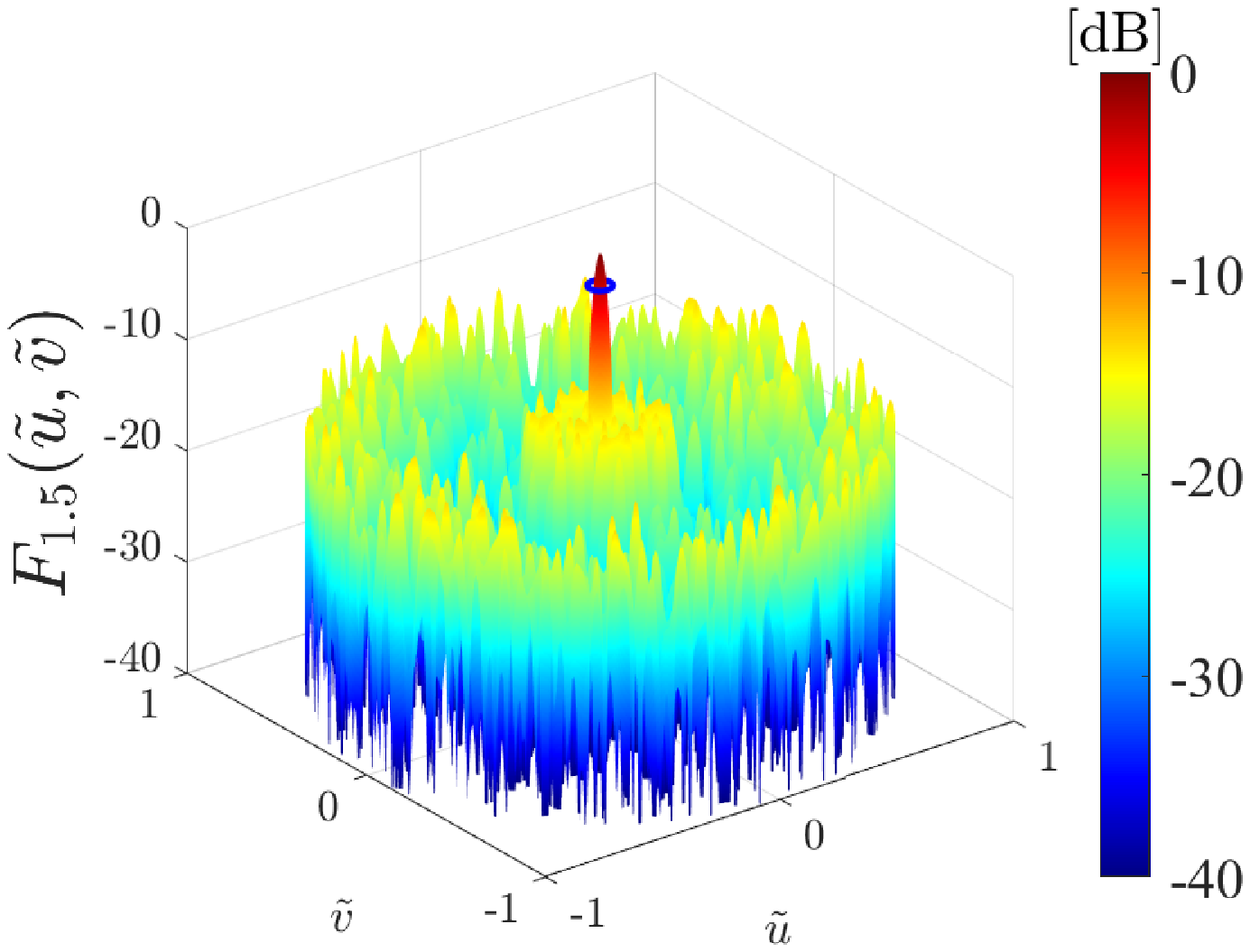}}
	\centering
	\caption{(a) The final geometry of subarrays where the blue squares are the centers of domino tiles, (b) maximum  SLL during iterative optimization on  the position of subarrays, (c) $d^{t_i,t_j}_s$,  the distance of tiles $t_i$ and $t_j$ when they are not in the same subarray and $x$-axis is the enumeration of $\{t_i,t_j\}$ pairs  and (d) the EBP of aperture with the blue circle on main beam indicating the 3dB-beamwidth.}
	\label{final_geo}
\end{figure}

Let us consider a uniform array for the JCAS aperture with $N_x\times N_y=30\times30$, with $\lambda/2$ inter-element distance where $\lambda$ is the wavelength at the center frequency of operation $f_c=28$ GHz. Firstly, these antennas are tiled with dominoes based on the maximum entropy, as  in Fig.~\ref{tiles}(a).
In order to show the capability of this large tiled array when scanning to $\theta_{min}=60^\circ$  according to 5G base station requirements \cite{ma2020thinned}, its EBP which corresponds to the scanning by linear beamforming is shown in Fig.~\ref{tiles}(b), where the maximum SLL is $-13.24$ dB and the beamwidth is $r_b=0.065$.
Then, the sunflower array is used to thin these dominoes. In order to cover the whole aperture, the distance between elements of the sunflower array is $s=3.5 \lambda$, while the angle steps of these elements are golden ratio  $\tau=1.618$.
The domino tiles which are in $r_c$ distance from the elements of sunflower are chosen for the subarrays and the thinning rate depends on this $r_c$. The simulation showed that $r_c=1.15\lambda_c$  provides the best performance for low SLL and minimum required tiles,  which results in overall 125 dominoes as shown in Fig.~\ref{SFA}. 
This elimination of dominoes increases maximum SLL, and reaches $-10.93$ dB, however the locations of subarrays can also be optimized for improving  the SLL
via iterative convex optimization (see appendix B).
The parameters in the position optimization of subarrays for an EBP with $\zeta=1.5$ are  $\tilde{r}_b=0.043$ for  the main beamwidth and  $\mu=\lambda/25$ for the maximum of position steps. The angles $0^\circ\leq\theta\leq90^\circ$ and $-180^\circ\leq\phi<180^\circ$ are swept with $1^\circ$. The objective function is to minimize the maximum SLL in the EBP.
As is shown in Fig.~\ref{final_geo}(a)-(b), the optimization reaches a maximum  SLL of $-13.95$ dB. For mutual coupling and implementation considerations, the inter-distances of each tile in a subarray with the adjacent tiles in other subarrays are checked to be larger than $\lambda$ which is shown in Fig.~\ref{final_geo}(c).  The EBP of the final geometry, has an SLL of  $-13.5$ dB (due to the linear approximation in the optimization) while having a beamwidth of $\tilde{r}=0.043$ as it is shown in Fig.~\ref{final_geo}(d). 
This final array geometry in Fig.~\ref{final_geo}(a) will be employed as the JCAS aperture.

\subsubsection{Robustness of STPA to wideband operation}
\begin{figure}[H]
    \centering
	\includegraphics[width=0.5\linewidth]{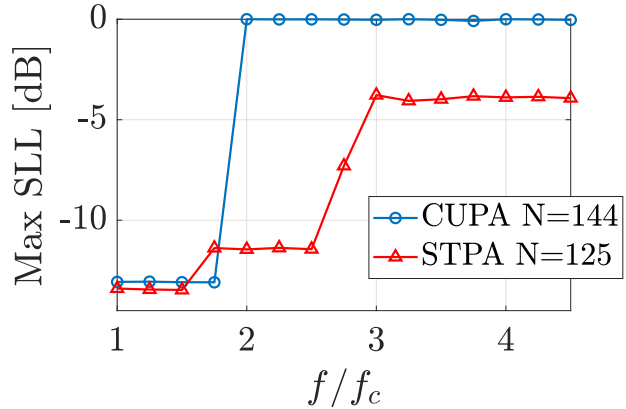}
	\caption{ The level of maximum SLL in EBP with $\zeta=2$  for  STPA and CUPA with 144 antennas.}
	\label{Ex56}
\end{figure}

 The array factor is the radiated field of the array when antennas are isotropic, and is critical for wideband operation.
Fig.~\ref{Ex56} demonstrates how this designed aperture performs with an increase in the frequency of operation. Where the EBP of STPA  at $f/f_c=2.5$  still shows  a maximum SLL of -11.44 dB, the CUPA with 144 antennas suffers from high SLL (grating lobes).  Considering 1.75$f_c$ and 2.5$f_c$ as maximum frequencies for CUPA and STPA, respectively, STPA contributes to 42\% improvement in maximum operational frequency of the array.
\subsubsection{Practical STPA design}
\begin{figure}[H]
	\subfigure[]
	{\includegraphics[width=0.315\linewidth]{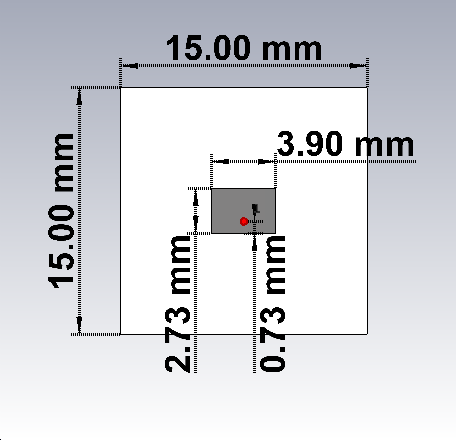}}\hspace{0.5cm}
	\subfigure[]
	{\includegraphics[width=0.35\linewidth]{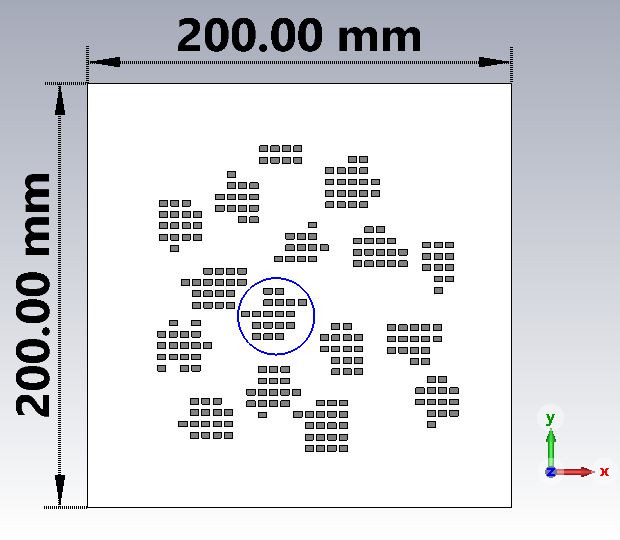}}
	\subfigure[]
	{\includegraphics[width=0.35\linewidth]{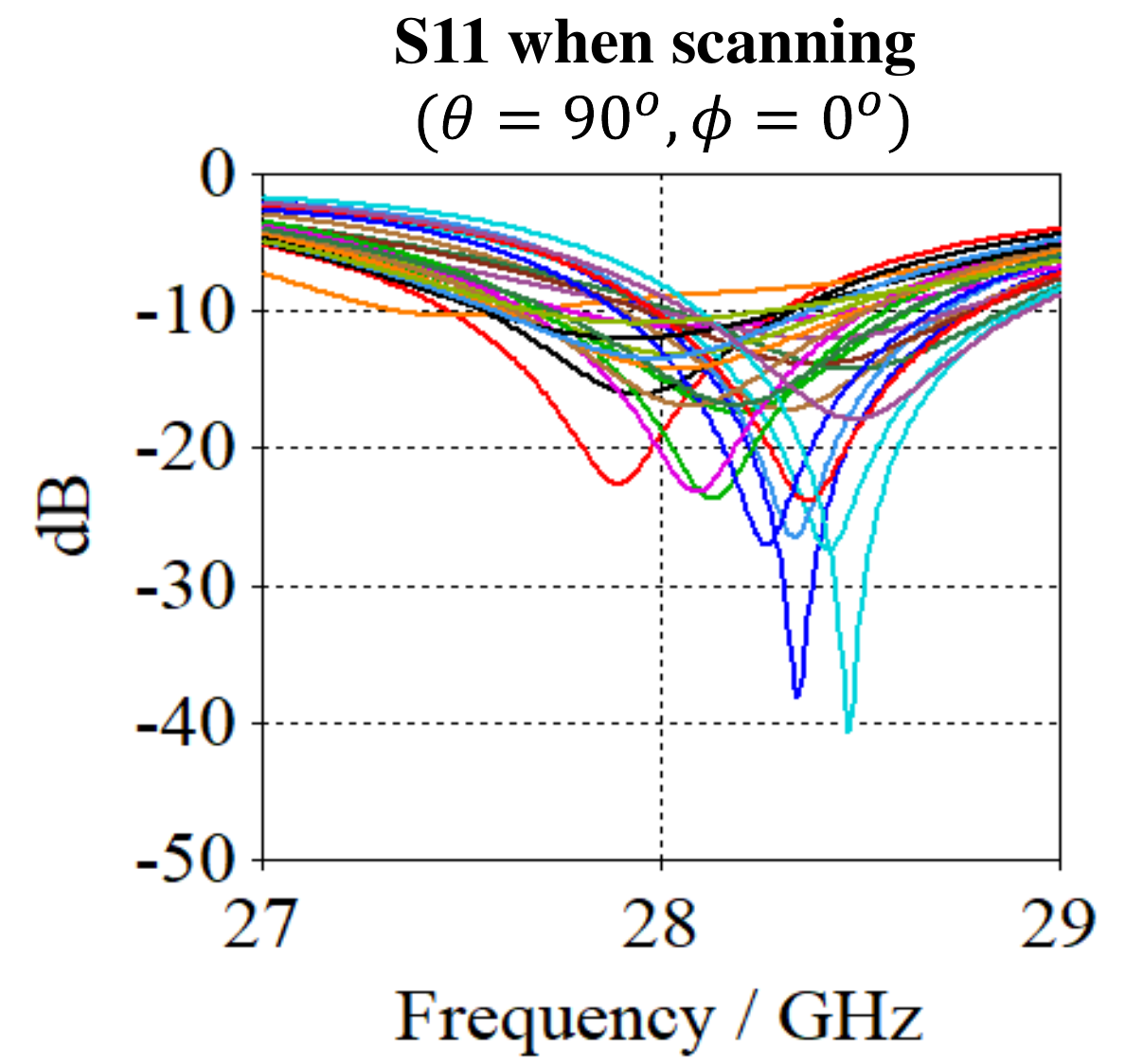}}
	\subfigure[]
	{\includegraphics[width=0.35\linewidth]{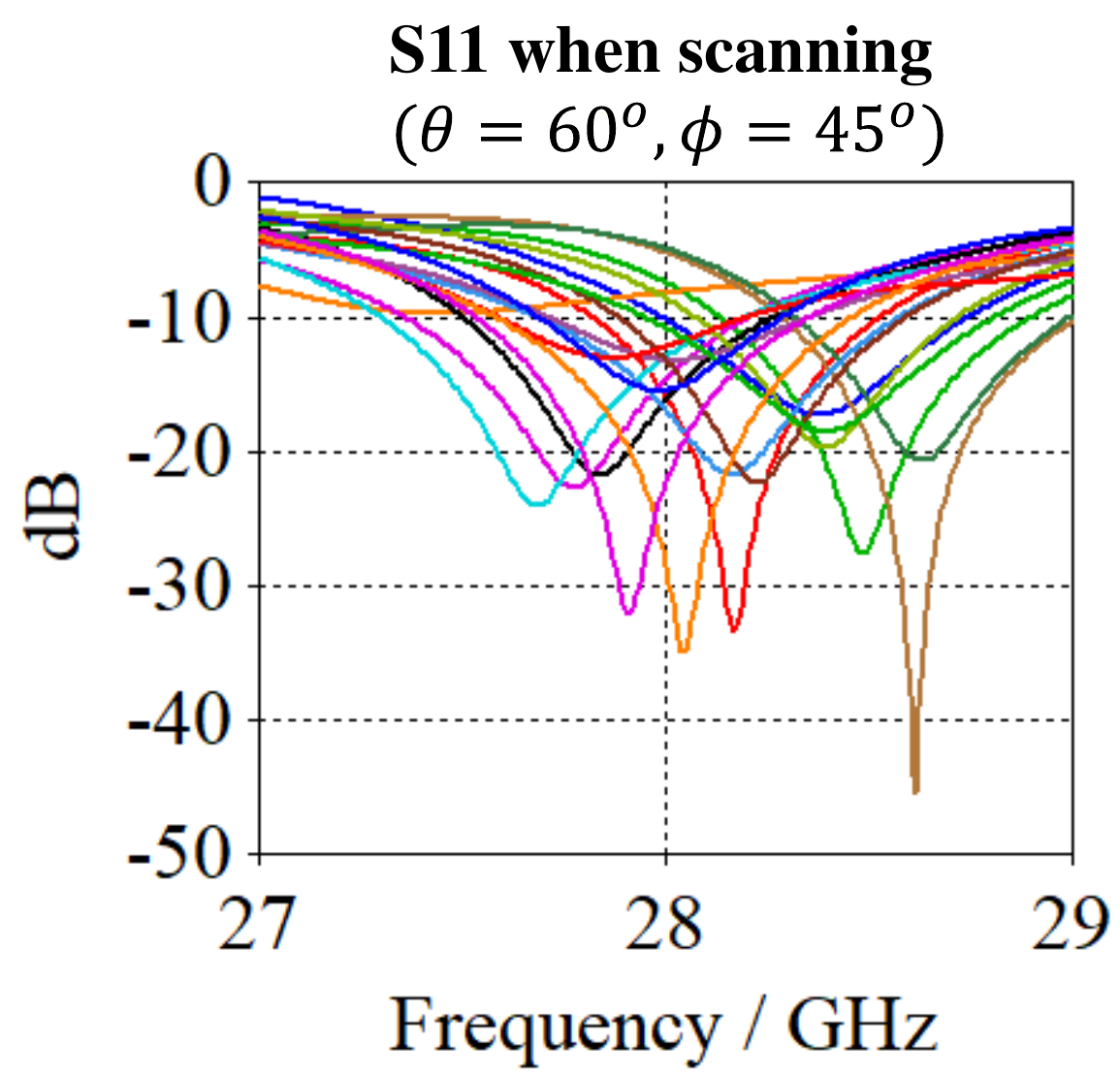}}
	\subfigure[]
	{\includegraphics[width=0.35\linewidth]{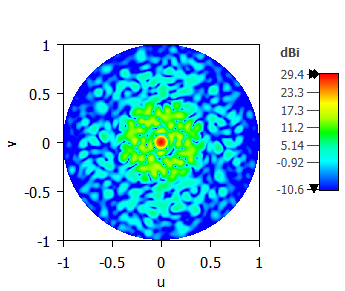}}
	\centering
	\subfigure[]
	{\includegraphics[width=0.35\linewidth]{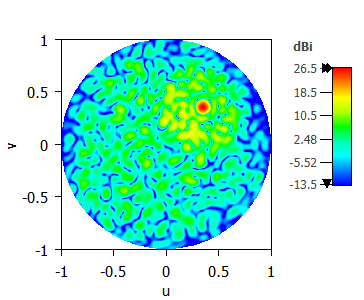}}
	\caption{(a) The geometry of the patch antenna element in CST, with the red dot denoting the feeding point (b) the arrays of antennas and the blue circle denoting subarray 1, (c) S11-parameters of 18 antennas at subarray 1 when scanning $(\theta=90^\circ,\phi=0^\circ)$, (d) S11-parameters of 18 antennas at subarray 1 when scanning $(\theta=60^\circ,\phi=45^\circ)$  (e) the  radiation pattern of the array when scanning $(\theta=90^\circ,\phi=0^\circ)$ (f) the  radiation pattern of array when scanning $(\theta=60^\circ,\phi=45^\circ)$.}
	\label{CST}
\end{figure}

An example of an STPA is simulated in CST Studio. The radiating elements are patch antennas as in Fig.~\ref{CST}(a) and designed  on the Rogers RO3003 substrate with thickness of $0.25$ mm, relative permittivity of $\epsilon_r=3$, copper cladding thickness of 35 $\mu $m and excited by a discrete port. At $f_c=28$ GHz, the 10dB-bandwidth is $863$ MHz and 3dB-beamwidth is $89.3^{\circ}<\theta_\text{3~dB}<119.1^{\circ}$ for various $\phi$s. Upon the knowledge of the positions of phase centers for tiles as shown in Fig.~\ref{final_geo}(a) and the orientation of dominoes (horizontal or vertical), the STPA has obtained and shown in  Fig.~\ref{CST}(b).
The radiation from the STPA  can be simulated by simultaneous  excitation  of the two horizontally or vertically juxtaposed patch antennas (corresponding to a tile) with the same phase.
The number of patch antennas is 250 whereas the number of phase shifters and tiles is 125.  As an example, the
 S11 parameters of 18 antennas (9 dominoes)  at subarray 1 are shown in Fig.~\ref{CST}(c)  when all antennas are excited simultaneously for scanning $(\theta,\phi)=(90^\circ,0^\circ)$ and Fig.~\ref{CST} (d) for scanning $(\theta,\phi)=(60^\circ,45^\circ)$. The changes in the resonance frequency and magnitude of S11 are noticeable due to the mutual coupling effects. 
The radiation from the STPA at the frequency of $28$ GHz and beamformed to boresight $(\theta,\phi)=(90^\circ,0^\circ)$  is shown in Fig.~\ref{CST}(e) where the maximum directivity is $29.4$ dBi,  the 3dB-beamwidth is $4^\circ$ and the maximum SLL is $-13.6$ dB, whereas in  case of beamformed towards  $(\theta,\phi)=(60^\circ,45^\circ)$,  the maximum directivity is $26.5$ dBi, the maximum SLL reaches $-11.9$ dB and beamwidth to $4.6^\circ$ as shown in Fig.~\ref{CST}(f). The decrease of directivity and SLL suppression in  the full-wave simulations is due to the non-isotropic radiation of patch antennas, the mutual coupling effects and the asymmetric radiations of horizontal and vertical tiles. 
\subsection{Performance metrics}
Once the practicality of STPA is demonstrated by full-wave simulation, we will proceed with simple isotropic modelling of radiating elements for analyzing the performance  for the communication and sensing in comparison with CUPA.
\subsubsection{CP without overlapping communication and sensing beams}
\begin{figure}[h!]
    \centering
    \includegraphics[width=0.4\linewidth]{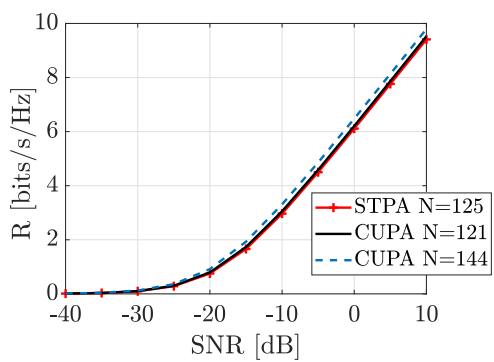}
    \caption{The spectral efficiency for the designed STPA with $N=125$ and CUPA with $N=121$ and $N=144$ when there is no overlapping of communication and sensing beams. SNR denotes the SNR level in the LOS path.}
    \label{Spectral}
\end{figure}

We consider the situation Fig.~\ref{CP}(a) and assume a user is at the range of $r_{UE}=50$ m and  the angle of $(\theta_{UE},\phi_{UE})=(70^\circ,50^\circ)$.
The channel in \eqref{Channel} from  the JCAS aperture to the user is modeled by a dominant LOS path, besides NLOS paths which include  $N_{cl}=8$ clusters each with  $N_{ray}=10$ rays and having 10 dB power less than the LOS path. The overall power of the channel, including LOS and NLOS path, is normalized for the channel as in \cite{ma2020thinned}. The scatters are distributed uniformly within all elevations $0^\circ\leq \theta\leq90^\circ$ and azimuth $-180^\circ\leq \phi<180^\circ$ angles. The channel from BS to the user via the target $\mathbf{H}_{target,1}$ is also considered in the NLOS path.
Based on the estimated channel and desired scanning angles, $\mathbf{w}_{JCAS}$ and based on that, the spectral efficiency $R$ can be calculated. The average of $R$ from 200 channel realizations is obtained for  the designed STPA. We intend to compare the performance of STPA with a symmetrical CUPA, so the closest numbers of antennas of CUPA are $N=11\times11$ and $N=12\times12$. As it is shown in Fig.~\ref{Spectral}, all three antenna configurations (STPA and 2 times CUPA) have similar performances,  where CUPA with $N=121$ and $N=144$  has slightly better spectral efficiency due to their higher directivity.
\begin{figure}[h!]
    \centering
    	\subfigure[]
{	\includegraphics[width=0.4\linewidth]{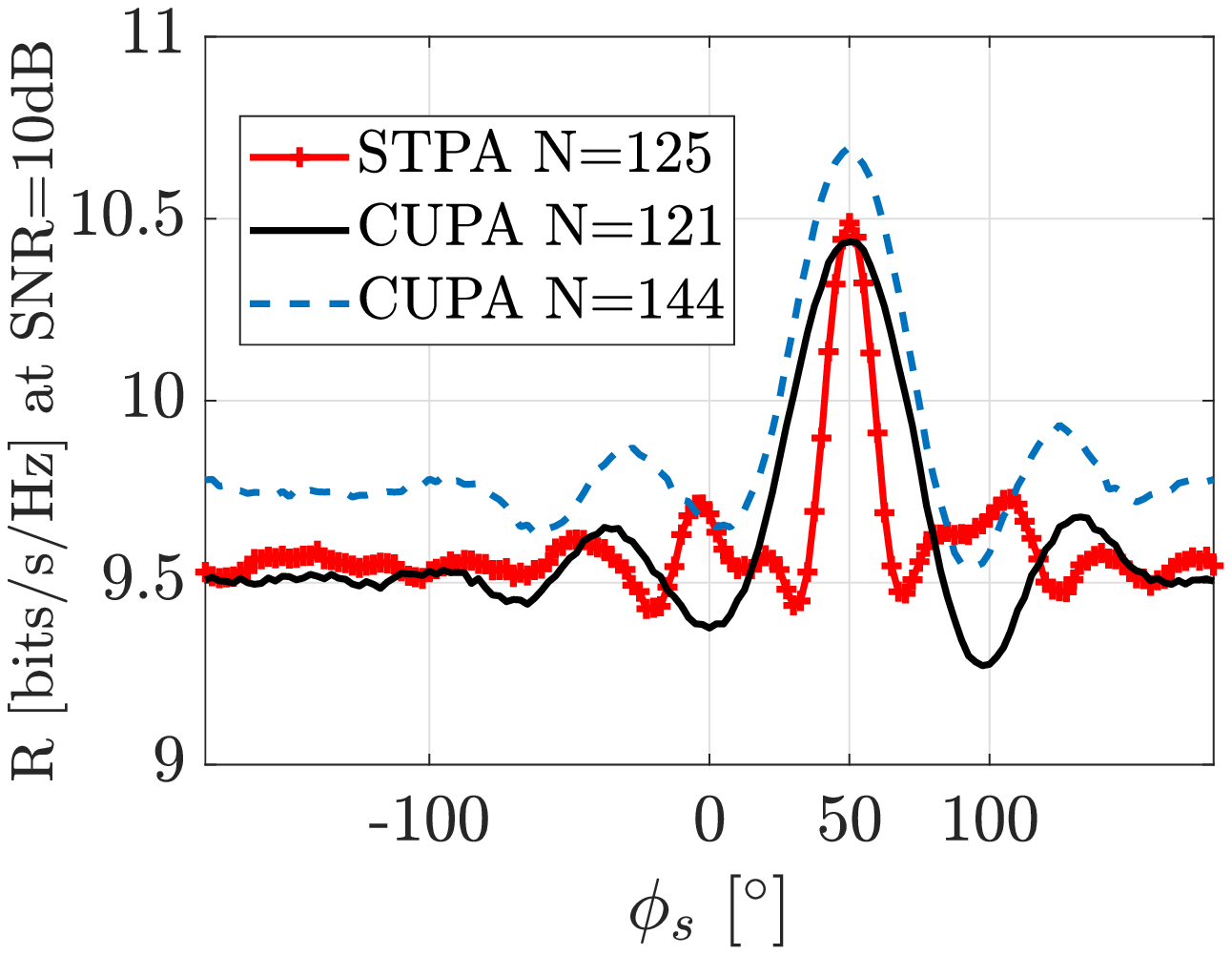}}
\subfigure[]
{	\includegraphics[width=0.4\linewidth]{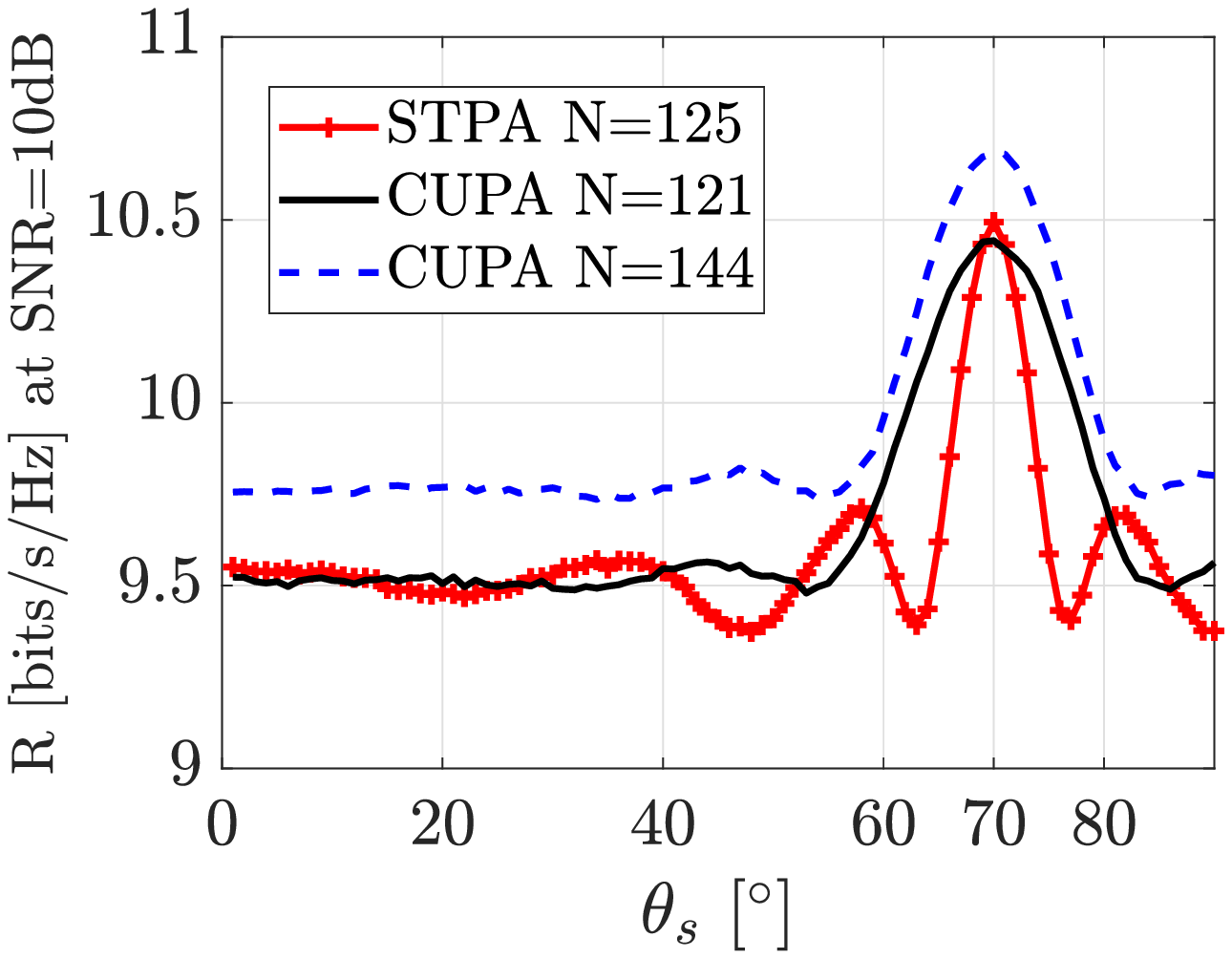}}
	\caption{The effect of the scanning beam  on the spectral efficiency for the STPA compared with CUPAs while scanning in (a) azimuth, and  (b) elevation.}
	\label{sc}
\end{figure}
\subsubsection{CP with overlapping communication
 and sensing beams when the UE blocks the sensing link}
As the same waveform is used for both beams, when overlapping occurs either in the scanning mode, or  in the tracking mode with the range of the tracking target being larger than the range of the UE, which is the case in Fig.~\ref{CP}(b), the signal power at the UE is increased, therefore spectral efficiency is supposed to increase.  Let us consider  the user is at the angle $(\theta_{UE},\phi_{UE})=(70^\circ,50^\circ)$, the sensing is in the scanning mode and  the sensing beam first scans over azimuth at the elevation $\theta=70^\circ$ and then over elevation at azimuth $\phi=50^\circ$ angles. The average spectral efficiencies at an SNR of 10 dB is presented in Fig.~\ref{sc}(a) and (b). The deviation from  the  spectral efficiency without overlapping  is noticeable for the STPA only when the difference of communication angle and scanning angle is $10^\circ$, whereas this number for the CUPA with $144$ and $121$ antennas is $25^\circ$. 
The Fig.~\ref{sc}(a) and (b) and above discussion are also valid for the  tracking mode of a target with larger range than the UE, e.g., $r_{UE}=70$ m and $r_{t}=140$ m. Regarding sensing performance in these cases, the received signal at the BS from the scanning beam when it reaches the angle of the UE, is zero due to zero reflection.
In the overlapping when the tracked target is behind the UE,  the received reflected signal at the BS is again zero and the target is also lost. The spectral efficiency of the communication link increased since the sensing and communication beam have the same waveform,  if one intends to use separate waveforms in separate beams then beam overlapping is detrimental for the SNR at the user.

\subsubsection{CP with overlapping communication
 and sensing beams when the sensing target blocks the communication link} 
\begin{figure}[h!]
    \centering
{	\includegraphics[width=0.5\linewidth]{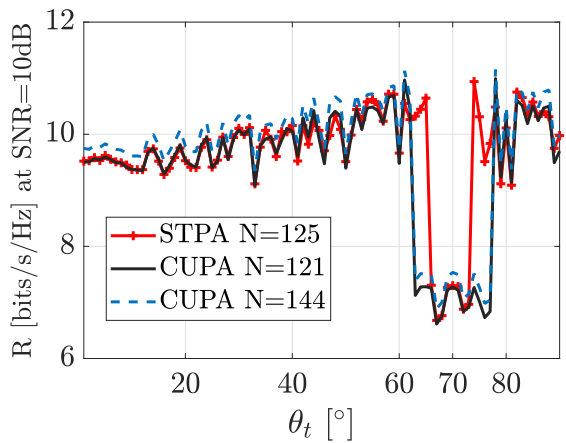}}
	\caption{The spectral efficiency when the target blocks the LOS path at the angle $\theta_t=70^\circ$ .}
	\label{OS}
\end{figure}

The range for UE is considered to be $r_{UE}=70$ m, the target body moves  at the constant azimuth of $\phi_t=50^\circ$, perpendicular to the LOS path and crossing it at $r_m=35$ m  as in Fig.~\ref{CP}(c).  The parameters for the body in \eqref{Gamma_t1} are assumed as $\epsilon_b=0.1-j2.33$, $\sigma_b=0.45$ dB \cite{bhardwaj2021geometrical} and the incident angle on the  body $\delta=\theta_t$. The variation of spectral efficiency at an SNR of 10 dB for the STPA and CUPAs  are shown in Fig.~\ref{OS}. 
There is an increased oscillation in the spectral efficiency just before and after blockage occurrence, which is due to constructive addition of $\mathbf{H}_{target,1}$ and $\mathbf{H}_{LOS}$ \cite{wu2022blockage}.
The most noticeable difference between the STPA and CUPAs is the time of blockage where the STPA, due to its narrower beamwidth, has shorter time. In these scenarios, the blockage time is proportional to the overlapping angles of communication and sensing beam, hence it is observed in Fig.~\ref{OS} that the STPA has a 40$\%$  shorter blockage time than CUPAs in the elevation sensing. 
\begin{figure}[h!]
    \centering
    \includegraphics[width=0.45\linewidth]{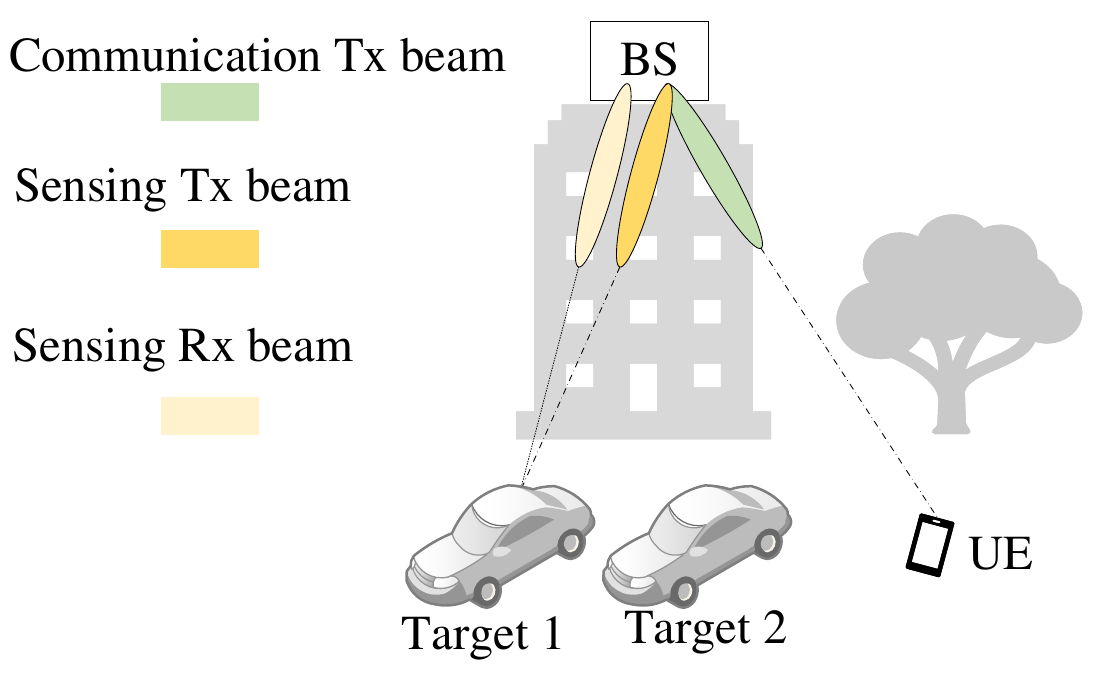}
    \caption{The scenario for sensing performance.}
    \label{S2target}
\end{figure}
\begin{table}[h!]
\centering
\caption{OFDM simulation parameters }
    \label{SF}
\begin{tabular}{ |c |c| c|}
\hline
 Parameter & Value & Description  \\ 
 \hline
 \hline
$f_c$ & 28 GHz & Carrier frequency \\ 
  \hline
$B$ & 40 MHz& Bandwidth \\
 \hline
$N_{sc}$ & 2048 &Number of subcarriers \\
 \hline
 $T_s$ & 51.2 $\mu$s & Symbol duration\\
 \hline
 $\Delta r$ & 3.75 m& Range resolution\\ 
 \hline
 $\Delta v$ & 3.07 m/s &Velocity resolution\\ 
 \hline
 $v_{umax}$& 52.32 m/s & Maximum unambiguous velocity\\
 \hline
  $r_{umax}$ & 7680 m& Maximum unambiguous range\\ 
 \hline
\end{tabular}
\end{table}
\subsubsection{Sensing performance of JCAS aperture}
 One of the key performance metric for sensing is the angular resolution in the scanning mode, which is studied for the STPA in comparison to CUPA with $N=144$ antennas.
Let us consider an outdoor scenario as in Fig.~\ref{S2target} where, in the worst case, two targets are located at the same ranges of $r_t=70$ m, velocities of $v_t=20$ m$/$s and the elevation of $\theta_t=70^\circ$  but in different azimuth angles. 
The OFDM simulation parameters are shown in the Table I.
The radiation gain of JCAS tiles and auxiliary antennas, the radar cross-section of targets and the noise figure of the receiver are assumed to be unit,  then the two-way channel gain in \eqref{Channel} is simply $b_l=\frac{\lambda^2}{(4\pi)^{3/2}r^2}$.  \par
\begin{figure}[H]
\centering
	\includegraphics[width=0.7\linewidth]{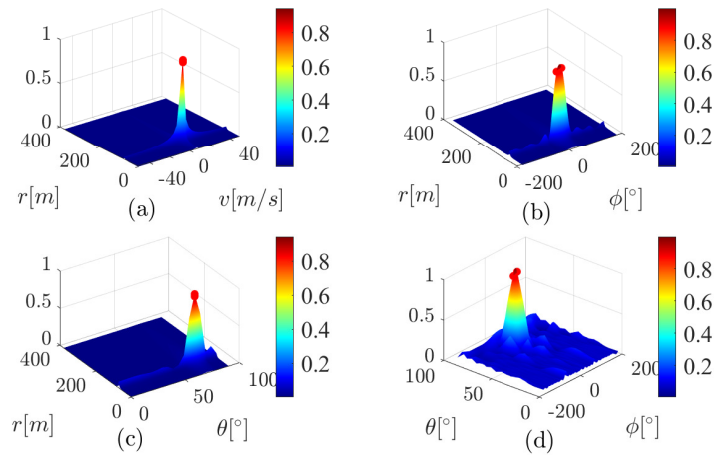}
	\caption{The normalized sensing images for CUPA with 144 antennas, where two targets and the communication user are at the same range, velocity, and elevation $\theta_t=70^\circ$ but
    at different azimuths
	$\phi_t=0^\circ$ and $20^\circ$. The two targets are \emph{not} distinguishable in $\phi$-axis. The red dots show the true feature of targets. }
	\label{Uniform144_sens}
\end{figure}
\begin{figure}[H]
\centering
	\includegraphics[width=0.7\linewidth]{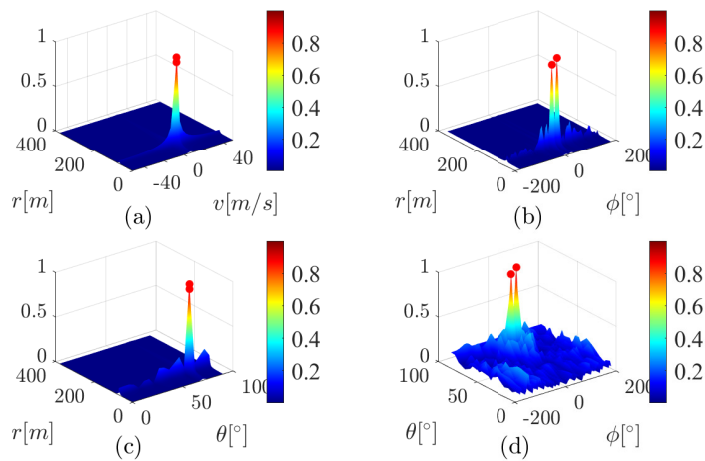}
	\caption{The normalized sensing images for the STPA, where two targets and the communication user are at the same range, velocity, and elevation $\theta_t=70^\circ$ but
    at different azimuths
	$\phi_t=0^\circ$ and $20^\circ$. The two targets are distinguishable in $\phi$-axis. The red dots show the true feature of targets. }
	\label{Tiled_Sens}
\end{figure}

The coherent processing duration for Doppler estimation is $N_dT_s$ where $N_d=34$. A communication packet includes $N_dN_e$ symbols, where $N_e=19$ is the number of elevation angles and scanning in azimuth is performed during $N_a=73$ communication packets, hence the elevation  and azimuth angles are swept by $5^\circ$. The  variance of noise $\tilde{z}_n$ in \eqref{y} is considered to be 10 dB lower than the received target signal.  As the emphasis here is on the angular resolution of JCAS aperture, we assume the angle differences of targets  are $20^\circ$ only in azimuth and compare the angle estimation of targets between the STPA and the CUPA with $N=144$. The range, velocity and elevation angle of targets are shown in Fig.~\ref{Uniform144_sens}(a) and (c) for the CUPA, and in Fig.~\ref{Tiled_Sens}(a) and (c) for the STPA. Both arrays provide a true estimation of these features of targets. However,
comparing the results in Fig.~\ref{Uniform144_sens}(b) and (d)  and Fig.~\ref{Tiled_Sens}(b) and (d)  show that the designed STPA is capable of distinguishing two adjacent targets in $\phi_t=0^\circ$ and $20^\circ$ where the CUPA with more number of antennas cannot. Therefore, the STPA provides a high angular resolution image during the scanning mode of sensing.

\subsection{Discussion}
In this paper, we assumed a scenario with a dominant LOS path, therefore the increase in the average SLL due to tiling and sparse design is not detrimental. The effect of interference on the JCAS performance of the STPA for NLOS case is still an open question.
As future work, the aperture efficiency of the designed STPA should also be improved. This can be tackled by assigning tiles with different sizes to outer subarrays. Besides, this study can be extended to hybrid beamforming arrays combined with phased-MIMO radar concepts.
\section{Conclusion}
Analog beamforming arrays can increase the gain of radiation and overcome propagation attenuation in millimeter wave communication and sensing links, therefore in alliance with the fully-connected analog beamforming arrays,
 a tiled sparse planar array (STPA) is proposed with modular design, narrow beamwidth and decreased  number of phase shifters for joint communication and sensing applications. The design is based on maximum entropy in the phase centers of the tiles in the array and sparse subarrays.  The results show that an STPA with 125 domino tiles has stable radiation  up to 2.5 times the center frequency of operation, $f_c$, whereas the maximum frequency of operation is $1.75f_c$ for CUPA with 144 antennas.  A realistic radiation pattern of STPA, including antenna imperfections, is shown by full-wave simulation. When the sensing and communication beams are not overlapping, CUPA and STPA guarantee the same spectral efficiency in the communication link, whereas
 a better angular resolution in sensing is achieved by the STPA, e.g., two targets in the  proximity of boresight are distinguished while their azimuth is 20$^\circ$ different.
 The analysis also includes the overlapping of communication and sensing beams and a blockage scenario by a tracked target. As the STPA has a narrower beamwidth than the CUPA, the blockage time is  40$\%$ shorter while the tracked target moves in the elevation angles. 

\section{Acknowledgment}
This work has received funding from the Dutch Sector Plan and the MSCA-IF-2020 V.I.P. project No. 101026885.

\section{Appendix}
\subsection{Sunflower array}
The sunflower array  can demonstrate the same  radiation pattern as a uniform array, but with a lower number of antennas. In polar coordinates, the $m$th element of the sunflower array  $(\rho_m,\psi_m)$ is obtained by \cite{vigano2011sunflower}:
\begin{equation}
    \rho_m=s\sqrt{\frac{m}{\pi}}, 
     \psi_m=2\pi m\tau \text{  where } m=1,2,...,M
\end{equation}
where $s$ is the inter-distance of elements, $\tau=1.618$ denotes the golden ratio and $M$ is the number of elements.
\subsection{ Position optimization}
The  positions of subarrays  are optimized based on an iterative convex optimization. Let us consider the expanded radiation field of the array in $(\tilde{u},\tilde{v})$-space as:
\begin{equation}
 f_{\zeta}(\tilde{u},\tilde{v})=\frac{1}{N}\sum_{n=1}^{N}e^{jk\zeta(\tilde{u}x_n+\tilde{v}y_n)},
  \label{rad00}
\end{equation}
 As it is discussed in  \cite{aslan2019multiple}, when  the positions of antennas are optimization variables, the problem is nonlinear and non-convex, however it can be solved  as a convex problem iteratively by first-order Taylor expansion as in:
\begin{equation}
\begin{split}
  &f_{\zeta}^{i}(\tilde{u},\tilde{v})=\frac{1}{N}\sum_{n=1}^{N} e^{jk\zeta\tilde{u}x_n^{i-1}}(1+jk\zeta\tilde{u} \epsilon_n^{i})\\&e^{jk\zeta
\tilde{v}y_n^{i-1}}(1+jk\zeta
\tilde{v} \beta_n^{i})\\& \approx \frac{1}{N}\sum_{n=1}^{N} e^{jk\zeta
\tilde{u}x_n^{i-1}}e^{jk\zeta
\tilde{v}y_n^{i-1}}\\&(1+jk\zeta
\tilde{u}
 \epsilon_n^{i}+jk\zeta
\tilde{v} \beta_n^{i}),\\&\text{ where }  
   |\beta_n^{i}|\ll 1, |\epsilon_n^{i}|\ll 1\\
  \end{split}
  \label{optim}
\end{equation}
where $i$ denotes the iteration index, and $\epsilon^{i}_n$ and $\beta^{i}$ 
 are the movement in $x$ and $y$ axes at $i$th iteration, respectively. $x^{i-1}_n$ and $y^{i-1}$ 
 are the position in $x$ and $y$ axes at $(i-1)$th iteration, respectively. \eqref{optim} is a linear
convex function that can be solved by Matlab CVX \cite{grant2013cvx}. To enforce only subarray movements, $\{\epsilon^{i}_n,\beta^{i}_n\}$s are equal for all antennas in a subarray.
Finally, the optimization can be written as:
\begin{equation}
\begin{aligned}
\min_{E^{i}, B^{i}} \quad & \gamma^i\\
\textrm{s.t.}
\quad&\max(|f_{\zeta}^{i}(\{\mathbf{\tilde{u},\tilde{v}}\}_{{\mathbf{SL}}})|)\leq \gamma^i, \\
 \quad&\text{real}(f_{\zeta}^{i}(0,0))=1,\\
 \quad& |\epsilon^{i}_n|\leq \mu,|\beta^{i}_n|\leq \mu, \hspace{0.7cm} \forall n \in \{1,2,...,N\}\\
\end{aligned}
\label{optformul}
\end{equation}
where $E^{i}=\{\epsilon^{i}_1,\epsilon^{i}_2,...,\epsilon^{i}_N\}$ and  $B^{i}=\{\beta^{i}_1,\beta^{i}_2,...,\beta^{i}_N\}$. $\{\mathbf{\tilde{u},\tilde{v}}\}_{{\mathbf{SL}}}$ is the set of all side lobe angles $(\tilde{u},\tilde{v})_\text{SL}$ and $\mu$ is a predetermined number. 
As the elements are domino tiles, there should be  a distance of at least $\lambda$ between the phase center of each domino at one subarray with the phase centers of other dominoes  at adjacent subarrays for fabrication possibility. This could also be added as a constraint to  \eqref{optformul} as discussed in \cite{aslan2019multiple}.

\bibliography{Ref}
\bibliographystyle{IEEEtran}
\end{document}